\documentclass[a4paper,11pt]{article}
\pdfoutput=1 

\usepackage{jheppub} 
\usepackage[T1]{fontenc} 
\usepackage{slashed}
\usepackage{graphicx,color}

\usepackage{dcolumn}
\usepackage{bm}
\usepackage{subfig}
\usepackage{graphicx}
\usepackage{amssymb}
\usepackage{amsmath}
\usepackage{stackrel}
\usepackage{pgfplots}
\usepackage{tikz-feynman}
\tikzfeynmanset{compat=1.0.0}
\usetikzlibrary{positioning}
\usetikzlibrary{snakes}

\title{\boldmath Probing Bottom-flavored Scalar Dark Matters at Loop Level}
\author[a]{Wei Chao,}
\author[a]{Jian-guo Jiang,}
\author[a]{Min Su}

\affiliation[a]{Center for advanced quantum studies, and Department of Physics,
Beijing Normal University, \\ Beijing 100875,  China}

\emailAdd{chaowei@bnu.edu.cn}
\emailAdd{jgjiang@mail.bnu.edu.cn}
\emailAdd{sumin@mail.bnu.edu.cn}

\abstract{In this paper we consider loop corrections to the spin-independent WIMP-nucleon scattering cross section in  bottom-quark flavored scalar-type dark matter models.  We focus on two scenarios: (a) a complex scalar dark matter with a scalar particle as the mediator; and (b) a real scalar dark matter with a vector boson as the mediator. In both scenarios,  the direct detection cross sections are either spin-dependent or kinematically forbidden at the tree-level.  Corrections induced by  the WIMP-gluon effective operator, scalar-type WIMP-quark effective operator, and the twist-2 effective operator are calculated. Numerical results show that loop induced spin-independent WIMP-nucleon scattering cross sections are quite considerable in both scenarios.}

\begin{document} 
\maketitle
\flushbottom

\section{Introduction}

Various astrophysical observations have confirmed the existence of cold dark matter (DM)~\cite{Aghanim:2018eyx}. However, what is DM made by and how it couples to the Standard Model (SM) particles still elude us.  During the past decades many DM candidates with masses ranging from $10^{-22}~{\rm eV}$ to $10^{55}~{\rm GeV}$ have been proposed, of which the weakly interacting massive particles (WIMPs)~\cite{Goldberg:1983nd,Ellis:1983ew,Jungman:1995df,Servant:2002aq,Cheng:2002ej,Bertone:2004pz}  are most promising as they can naturally explain the observed relic density with their masses at the electroweak scale and their interactions as weak as the weak nuclear force.  

There are many experiments~\cite{Lin:2019uvt,Slatyer:2017sev,Hooper:2018kfv} on the Earth designed for probing WIMPs, which can be classified into two categories: the direct detection experiment and the indirect detection experiment.  Direct detection experiments measure the  nuclear (or the electron) recoil energy induced by the elastic scattering of WIMPs with nucleon (electron) in underground laboratories. Indirect detection experiments detect the flux of secondary cosmic rays injected by the WIMP annihilations or decays. These two methods are potentially complimentary to each other in testing a WIMP model.

Theoretically, it will be a good strategy to investigate the direct detection signal of a WIMP model that may have non-trivial signals in indirect detection experiments. 
The AMS-02 collaboration~\cite{Aguilar:2016kjl} has identified an excess of cosmic-ray antiprotons, which can be explained by the DM annihilation into $b\bar b$ with a reduced cross section of $(0.8\sim5.2) \times 10^{-26}~{\rm cm^3/s}$ for a DM mass around $(64,~88)~{\rm GeV}$~\cite{Cholis:2019ejx}.  
Besides, there is an excess of GeV-scale gamma-rays~\cite{Goodenough:2009gk,Hooper:2010mq,Hooper:2011ti,Abazajian:2012pn,Daylan:2014rsa,Hooper:2019xss} observed  from the region surrounding the Galactic Center. The spectral shape, morphology and intensity  of the excess can also be explained by the annihilation of DM into $b\bar b$ with the mass in the range of $(40,~70)~{\rm GeV}$~\cite{Berlin:2014tja}.  
Motivated by the fact that the bottom-quark flavored DM  models may have non-trivial signals in indirect detection experiments, we study signals of these models in direct detection experiments. 
It should be noted that  the antiproton and GeV gamma ray excesses can also be interpreted by astrophysical sources, and we are not trying to explain these excesses by a  bottom-flavored DM model.
Instead, our purpose is to provide loop-level analytical result of DM-nucleon scattering cross section that  may help to identify a bottom-quark flavored DM in future combined searches of direct and indirect detection experiments.

Due to the technological innovations and advances,  the precision and detecting efficiency of direct detection experiments have been greatly improved.  For WIMP models,  whose scattering cross section with nucleon is spin-independent at the tree-level, the existing experimental accuracy is already able to detect their sensitive parameter space. In this paper, we will focus on two interesting scenarios: (A) a complex scalar DM with a scalar particle as the mediator and (B) a real scalar DM with a spin-1 vector boson as the mediator. In both scenarios, the DM-nucleon cross section is either spin-dependent or kinematically forbidden at the tree-level. As a result, loop corrections~\cite{Haisch:2013uaa,Crivellin:2014gpa,DEramo:2016gos,Crivellin:2014qxa,Bishara:2018vix,Li:2018qip,Sanderson:2018lmj,Hisano:2010fy,Hisano:2011cs,Ertas:2019dew,Ishiwata:2018sdi,Abe:2018emu,Chao:2018xwz,Ghorbani:2018pjh,Li:2019fnn,Chao:2019lhb} turn to be the dominant contribution to the spin-independent DM-nucleon cross section. Since the exclusion limit given by direct detections will soon reach the so-called neutrino floor~\cite{Monroe:2007xp,Strigari:2009bq,Billard:2013qya,Gelmini:2018ogy,Boehm:2018sux,Chao:2019pyh}, below which the DM signal is indistinguishable from an irreducible background induced by the coherent elastic neutrino-nuclei scattering, loop-level calculations will be necessary whenever one wants to find out  whether or not such a model is detectable using the current direct detection techniques. Following Refs.~\cite{Hisano:2015bma,Hisano:2017jmz}, we calculate Wilson coefficients of the scalar-type and twist-2 DM-bottom-quark effective operators which arise from the box and triangle diagrams, as well as Wilson coefficient of DM-gluon effective operator~\cite{Hisano:2010ct,Hisano:2015rsa} which arises at the two-loop level.  After accounting for constraints of the observed relic abundance, we find that 
\begin{itemize}
\item Loop corrections are significant in the scenario A, and most of the parameter spaces of interest to this model can be tested using current direct detection techniques, as shown in the left-panel of the Fig.~\ref{fig:cross section}.
\item For scenario B, loop effects are relatively small and  the DM-nucleon scattering cross section lies above the neutrino floor only in the low DM mass regime, as shown in the right-panel of the Fig.~\ref{fig:cross section}. 
\end{itemize}
It should be mentioned that our analytical results can be directly applied to study the parameter space of interest in future indirect detection experiments. 


The remaining of this paper is organized as follows: In section II we calculate the corrections to the DM-nucleon cross section in the scenario A. Section III is devoted to study the direct detection cross section of the scenario B.  Numerical results are given in the section IV and the last part is conclusion.  The thermal average of the reduced annihilation cross sections, loop integrals and nucleon form factors are given in the appendices,~\ref{annihilation}, \ref{loop} and \ref{nucleon matrix elements}.

\section{Complex scalar DM with spin-0 mediator}

Since we are going to calculate the DM-nucleon scattering cross section at the loop level, a description of DM interactions in the effective field theory approach  does not apply.
Throughout this paper we consider the direct detection signals of a scalar dark matter, whose interactions take a general form: ${\cal L} \supset  \overline{\rm DM} {\rm DM } ~{\rm Med}+ \overline{\rm SM} {\rm SM} ~{\rm Med}$, with ${\rm Med}$ stands for the mediator particle. 
The Lagrangian of a complex scalar dark matter $\varphi$, that couples to the bottom quark via a spin-0 mediator $\Phi$, can be written as
\begin{equation}
\label{eq:scalardm}
\begin{split}
-{\cal L}\sim {1\over 2} \lambda \varphi^\dagger\varphi \Phi^2 + \Lambda \varphi^\dagger \varphi  \Phi + f_S \Phi \bar b b + f_P^{} \Phi \bar b i\gamma_5^{} b + {\rm h.c.}
\end{split}
\end{equation}
where $\Lambda$ is the coupling with mass dimension. $\Lambda$ and $\lambda$ are taken as free parameters, actually they might be correlated with each other via a Higgs mechanism, i.e., $\Lambda= \lambda v_\Phi$ with $v_\Phi$ the vacuum expectation value of the $\Phi$. Eq.~\eqref{eq:scalardm} is invariant under a $Z_2$ discrete flavor symmetry $\varphi \leftrightarrow -\varphi$, which stabilizes the $\varphi$ as a DM candidate. If there is another $Z_2$ symmetry for $\Phi$, i.e., $ \Phi\to -\Phi$,  then the term $\Lambda \varphi^\dagger \varphi \Phi + {\rm h.c.}$ will be forbidden. The Yukawa interactions in Eq.~\eqref{eq:scalardm} may come from integrating out a new vector-like heavy fermion, $\psi$, which couples to the third generation quark doublet as well as the right-handed bottom quark, $\tilde{y}_b^{} \overline{Q_L^{3}} H \psi_R^{} + \zeta \overline{\psi_L^{} } \Phi b_R^{} + {\rm h.c.}$, where $H$ is the SM Higgs doublet, $\tilde{y}_b$ and $\zeta$ are Yukawa couplings. Apparently there is a rephrasing invariant term in the Lagrangian, $\arg(y_b\tilde{y}_b^* \zeta^* M_\psi)$, which can lead to the CP violation. In this paper, we will only focus on the pseudo-scalar interaction (only $f_P\neq 0$), as the scalar-type interaction will result in spin-independent scattering cross section at the leading order, which is already suppressed by the exclusion limits put by various dark matter direct detection experiments.

\subsection{Relic density}
The dark matter number density $n$, is governed by the Boltzmann equation~\cite{1991NuPhB.360..145G}:
\begin{eqnarray}
\dot{n} + 3Hn = - \langle \sigma v_{\rm M\slashed{o}ller} \rangle ( n^2 -n_{\rm EQ}^2 ) \; , \label{boltzmann}
\end{eqnarray}  
where $H $ is the Hubble constant, $\sigma v_{\rm M\slashed{o}ller}$ is the total annihilation cross section multiplied by the M$\slashed{\rm o}$ller velocity, $v_{\rm M\slashed{o}ller}=(|v_1 -v_2 |^2 -|v_1 \times v_2 |^2 )^{1/2}$, brackets denote thermal average and $n_{\rm EQ}$ is the number density at the thermal equilibrium. It has been shown that $\langle \sigma v_{\rm M\slashed{o}ller} \rangle =\langle \sigma v_{\rm lab} \rangle = 1/2 [1 + K_1^2 (x) /K_2^2 (x)] \langle \sigma v_{\rm cm} \rangle$~\cite{1991NuPhB.360..145G}, where $x=m/T$, $K_i$ is the modified Bessel functions of order $i$. 

\subsubsection{The $m_{\varphi}<m_{\Phi}$ scenario}

\begin{figure}
\begin{center}
\begin{tikzpicture}
\begin{feynman}
\vertex (i1);
\vertex [right=0.75cm of i1] (i2);
\vertex [right=1.5cm of i2] (i3);
\vertex [right=0.75cm of i3] (i4);
\vertex [below=1.5cm of i1] (i5);
\vertex [below=0.75cm of i2] (i6);
\vertex [below=0.75cm of i3] (i7);
\vertex [below=1.5cm of i4] (i8);
\diagram* {
	(i1) -- [scalar, ultra thick] (i6) -- [scalar, ultra thick] (i5),
	(i6) -- [scalar, ultra thick] (i7),
	(i8) -- [fermion, ultra thick] (i7) -- [fermion, ultra thick] (i4),
};
\end{feynman}
\node[red, thick] at (0.2,0.3) {$\varphi$};
\node[red, thick] at (0.2,-1.8) {$\varphi$};
\node[red, thick] at (1.5,-0.4) {$\Phi$};
\node[red, thick] at (2.8,0.3) {$b$};
\node[red, thick] at (2.8,-1.8) {$\bar{b}$};
\end{tikzpicture}
\hspace{1cm}
\begin{tikzpicture}
\begin{feynman}
\vertex (i1);
\vertex [below=1.5cm of i1] (i4);
\vertex [right=1.5cm of i1] (i2);
\vertex [right=1.5cm of i2] (i3);
\vertex [right=1.5cm of i4] (i5);
\vertex [right=1.5cm of i5] (i6);

\diagram* {
	(i1) -- [scalar, ultra thick] (i2) -- [scalar, ultra thick] (i3),
	(i4) -- [scalar, ultra thick] (i5) -- [scalar, ultra thick] (i6),
	(i2) -- [scalar, ultra thick] (i5)

};
\end{feynman}
\node[red, thick] at (0.2,0.3) {$\varphi$};
\node[red, thick] at (0.2,-1.8) {$\varphi$};
\node[red, thick] at (1.2,-0.75) {$\varphi$};
\node[red, thick] at (2.8,0.3) {$\Phi$};
\node[red, thick] at (2.8,-1.8) {$\Phi$};
\end{tikzpicture}
\hspace{1cm}
\begin{tikzpicture}
\begin{feynman}
\vertex (i1);
\vertex [below=1.5cm of i1] (i4);
\vertex [right=1.5cm of i1] (i2);
\vertex [right=1.5cm of i2] (i3);
\vertex [right=1.5cm of i4] (i5);
\vertex [right=1.5cm of i5] (i6);

\diagram* {
	(i1) -- [scalar, ultra thick] (i2) -- [scalar, ultra thick] (i6),
	(i4) -- [scalar, ultra thick] (i5) -- [scalar, ultra thick] (i3),
	(i2) -- [scalar, ultra thick] (i5)
};
\end{feynman}
\node[red, thick] at (0.2,0.3) {$\varphi$};
\node[red, thick] at (0.2,-1.8) {$\varphi$};
\node[red, thick] at (1.2,-0.75) {$\varphi$};
\node[red, thick] at (2.8,0.3) {$\Phi$};
\node[red, thick] at (2.8,-1.8) {$\Phi$};
\end{tikzpicture}
\hspace{1cm}
\begin{tikzpicture}
\begin{feynman}
\vertex (i1);
\vertex [right=0.75cm of i1] (i2);
\vertex [right=0.75cm of i2] (i3);
\vertex [below=1.5cm of i1] (i4);
\vertex [below=0.75cm of i2] (i5);
\vertex [below=1.5cm of i3] (i6);

\diagram* {
	(i1) -- [scalar, ultra thick] (i5) -- [scalar, ultra thick] (i3),
	(i4) -- [scalar, ultra thick] (i5) -- [scalar, ultra thick] (i6),
};
\end{feynman}
\node[red, thick] at (0,0.3) {$\varphi$};
\node[red, thick] at (0,-1.8) {$\varphi$};
\node[red, thick] at (1.5,0.3) {$\Phi$};
\node[red, thick] at (1.5,-1.8) {$\Phi$};
\end{tikzpicture}
\end{center}
\caption{Annihilation channels of complex scalar dark matter $\varphi$ with a spin-0 mediator $\Phi$. All Feynman diagrams are drawn with the help of TikZ-Feynman \cite{Ellis:2016jkw}.}
\label{fig:scalar annihilation}
\end{figure}
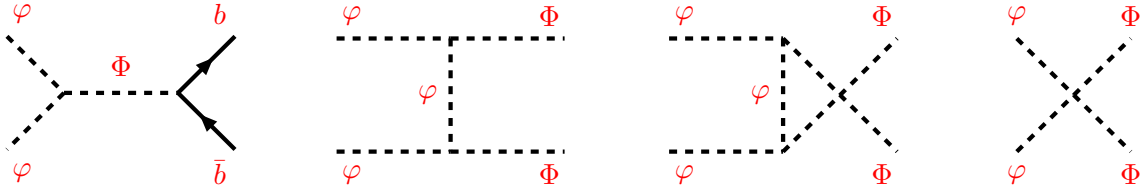

In the mass regime $m_\varphi<m_\Phi$, the annihilation channel $\varphi\varphi \to \Phi\Phi$ is kinematically forbidden and there is only one annihilation channel $\varphi\varphi \to \bar{b}b$ which is given in the  most left plot of the Fig.~\ref{fig:scalar annihilation}. The annihilation cross section can be written as 
\begin{equation}
\label{eq: anni-small}
\begin{split}
\sigma(\varphi \varphi \to \bar b b)  = {1\over 8\pi} \sqrt{{s-4m_b^2 \over s-4m_\varphi^2}} { \Lambda^2 f_P^2 \over (s-m_\Phi^2)^2 + m_\Phi^2 \Gamma_\Phi^2},
\end{split}
\end{equation}
where $\Gamma_\Phi$ is the decay rate of $\Phi$, 
\begin{equation}
\label{eq: decayrate}
\begin{split}
\Gamma_\Phi = {1 \over 8 \pi} f_P^2 \sqrt{m_\Phi^2-4m_b^2} + \Theta(m_\Phi -2 m_\varphi) {1 \over 16 \pi} {\Lambda^2 \over m_\Phi^2} \sqrt{m_\Phi^2 -4m_\varphi^2} ,
\end{split}
\end{equation}
with $\Theta(x)$ the unit step function. Analytically one can approximate the thermal average $\langle \sigma v\rangle$ with the non-relativistic expansion $\langle \sigma v \rangle = a + b\langle v^2 \rangle$ in the laboratory frame, where $a$ and $b$ are given in the Appendix \ref{annihilation}.

The present relic density of the DM is simply given by $\rho_\chi = m_\chi n_\chi = m_\chi s_0 Y_{\infty} $,  where $s_0$ is the present entropy density. The relic density can finally be expressed in terms of the critical density~\cite{Bertone:2004pz}
\begin{eqnarray}
\Omega h^2 \approx 2 \times {1.07 \times 10^9 {\rm GeV}^{-1}  x_F \over M_{pl} \sqrt{g_*} ( a + 3 b/x_F )} \; ,
\end{eqnarray}
where $M_{pl}$ is the Planck mass, $a$ and $b$, expressed in ${\rm GeV}^{-2}$, are the $s$-wave and the $p$-wave parts of the reduced annihilation cross section and $g_*$ is the effective degrees of freedom at the freeze-out temperature $T_F$, $x_F= M/T_F $, which is of the order ${\cal O}(22)$, the factor $2$ on the right-handed side accounts for the fact that the dark matter is a complex scalar. 
\subsubsection{The $m_{\varphi}>m_{\Phi}$ scenario}
In the mass regime $ m_\varphi >m_\Phi$, the annihilation channel $\varphi \varphi \to \Phi \Phi$, as shown in the remaining plots of the Fig.~\ref{fig:scalar annihilation}, is kinematically allowed. The annihilation cross section in the center-of-mass framework is 
\begin{eqnarray}
\sigma(\varphi \varphi \to \Phi \Phi) ={1 \over 32 \pi s} \sqrt{s-4m_\Phi^2 \over s-4m_\varphi^2} \left[ {2\Lambda^4 \over A^2 -B^2} + \frac{ 2\Lambda^2 (\Lambda^2 -2 A \lambda) }{AB}\tanh^{-1}\left({B \over A}\right)  + \lambda^2\right]
\label{varphiPhi}
\end{eqnarray} 
where $A=s/2-m_\Phi^2$ and $B=1/2\sqrt{(s-4m_\varphi^2)(s-4m_\Phi^2)} $. One may get the thermal average of the reduced annihilation cross section with the help of Eq.~(\ref{varphiPhi}), and expressions of $s$-wave and $p$-wave contributions are given in Appendix \ref{annihilation}, with the help of which one may estimate the relic abundance of $\varphi$.  

\subsection{Direct detection }
The dark matter direct detection experiments aim to observe phonon, light or ionization generated by the recoiled nuclei arising from the scattering of dark matter off the nuclei. The WIMP event rate can be written as~\cite{Lewin:1995rx}   
\begin{eqnarray}
{d R \over d E_R} = {MT} \times  {\rho_{\chi }\sigma^0_n A^2   \over  2 m_{\chi} \mu_n^2}   F^2 (E_R^{} ) \int_{v_{\rm min}}  {f(\vec{v}) \over v } d^3 v \label{mastereq}
\end{eqnarray}
where $M$ is the target mass, $T$ is the exposure time, $\rho_{\chi} $  is the DM density in the local halo, $\mu_n$ is the DM-nucleon reduced mass, $\sigma_n^0$ is the DM-nucleon cross section, $A$ is the atomic number, $F(E_R^{})$ is the nuclear form factor and we use the Helm form factor~\cite{Helm:1956zz}, $f(\vec{v})$ is the DM velocity distribution, $v_{\rm min}$ depends on $E_R$: $v_{\rm min} =\sqrt{m_T^{} E_R^{} /2\mu_T^2}$ with $\mu_T$ the DM-nucleus reduced mass. 

There are two ways measuring the local DM density $\rho_{\chi}$~\cite{Read:2014qva}: local measures that use the vertical kinematics of stars near the Sun, called tracers, and global measures that extrapolate $\rho_{\chi}$ from the rotation curve.   Its value is given with a large uncertainty as $\rho_\chi=(0.2-0.6 )~{\rm GeV}$.
There is no way to measure $f(\vec{v})$ directly,  it is usually assumed to be the Maxwell-Boltzmann distribution function   in the galactic center coordinate. The velocity integral in eq.~(\ref{mastereq}) can be analytically written as~\cite{Barger:2010gv}
\begin{eqnarray}
\int_{v_{\rm min}}  {f(\vec{v}) \over v } d^3 v  = {1\over 2 v_0^{}  \eta_E^{} } \left[{\rm erf} (\eta_+^{} ) - {\rm erf} (\eta_-^{} ) \right]- {1\over \pi v_0  \eta_E^{} }(\eta_+^{} -\eta_-^{} ) e^{-\eta_{\rm esc}^2}
\end{eqnarray}
where $v_0$ is the speed of the Local Standard of Rest, $\eta_E=v_E/v_0$ with $v_E$ the Earth velocity with respect to the galactic center, $\eta_{\rm esc} =v_{\rm esc} /v_0$ with $v_{\rm esc}$  the escape velocity of DM from our galaxy, $\eta_{\pm} ={\min} (v_{\rm min}/v_0 \pm \eta_E, v_{\rm esc}/v_0)$.  We take $v_0=220~{\rm km/s}$, $v_{\rm esc} =544~{\rm km/s}$
and $\vec{v}_E = \vec{v}_\odot +\vec{v}_\oplus \approx \vec{v}_\odot =232~{\rm km/s}$, where $\vec{v}_\odot$ and $\vec{v}_\oplus$ are the velocity of the Sun with respect to the Galaxy as well as the Earth rotational velocity, respectively. 

According to the Eq.~(\ref{mastereq}),  the non-observation of any DM signal  for a concrete DM direct detection experiment with fixed exposure,  will  generate the exclusion limit on the direct detection cross section in the $m_{\rm DM}-\sigma_n^0$ plane, so an analytical calculation of $\sigma_n^0$ is necessary so as to put constraint on the parameter space of the model.  For the complex scalar dark matter $\varphi$ with spin-0 mediator, the scattering cross section is spin-dependent and the effective operator at the quark level can be written as
\begin{eqnarray}
{\cal L}_b\sim -{ \Lambda f_P^{} \over m_\Phi^2 } \phi^\dagger \phi \bar b i\gamma_5^{} b 
\end{eqnarray}
Bottom quark is heavier than the proton and should be integrated out of an effective theory describing physics at nuclear scale.
Integrating out the bottom quark, one has loop induced coupling with gluon~\cite{Shifman:1978zn},
\begin{eqnarray}
{\cal L}_g\sim {\alpha_s \over 8\pi }{ \Lambda f_P^{} \over m_b^{}  m_\Phi^2 } \phi^\dagger \phi G_{\mu\nu
}^a \widetilde{G}^{a\mu\nu}
\end{eqnarray}
where $\widetilde{G}^{a}_{\mu\nu} =\varepsilon_{\mu\nu \rho \sigma}^{} G^a_{\rho\sigma}$ with the convention $\varepsilon_{0123} =1$. Now one can write down the effective Lagrangian at the nucleon level
\begin{eqnarray}
{\cal L}_N^{} \sim m_N \bar m \left(\sum_{q=u,d,s}{\Delta_q^N \over m_q } \right) {\Lambda f_P^{} \over  m_b^{} m_\Phi^2} \phi^\dagger\phi  \bar N i\gamma_5^{} N
\end{eqnarray}
where  $\bar m = (1/m_u+1/m_d+1/m_s)^{-1}$, $\Delta_u^{N} =0.84$, $\Delta_d^N=-0.43$ and $\Delta_s^N=-0.09$~\cite{Ellis:2008hf}.  The spin-dependent DM-nuclei cross section can be written as 
\begin{eqnarray}
\sigma_{SD}^{} = {1 \over 32\pi} {q^2 \over (m_\varphi+m_N)^2}\left[m_N \bar m \left(\sum_{q=u,d,s}{\Delta_q^N \over m_q } \right) {\Lambda f_P^{} \over  m_b^{} m_\Phi^2}\right]^2
\end{eqnarray}
where $q$ is the momentum transfer. Apparently this cross section is kinematically suppressed by the factor $q^2$. One needs to calculate the spin-independent cross section at the next-to-leading order to probe the model.

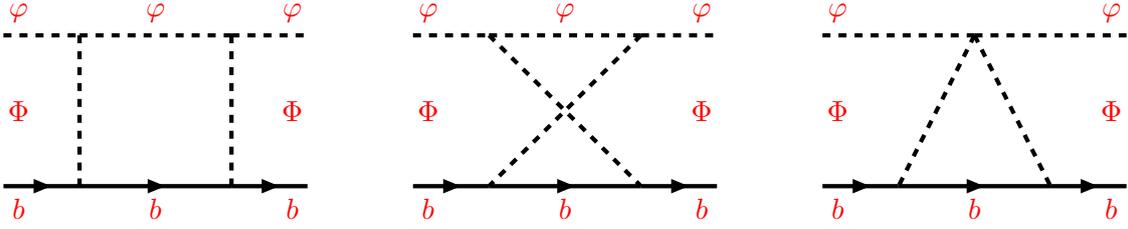
\begin{figure}
\begin{center}
\begin{tikzpicture}
\begin{feynman}
\vertex (i1);
\vertex [below=2cm of i1] (i2);
\vertex [right=1cm of i1] (i3);
\vertex [right=1cm of i2] (i4);
\vertex [right=2cm of i3] (i5);
\vertex [right=2cm of i4] (i6);
\vertex [right=1cm of i5] (i7);
\vertex [right=1cm of i6] (i8);
\diagram* {
	(i1) -- [scalar,ultra thick] (i3) -- [scalar, ultra thick] (i5) -- [scalar, ultra thick] (i7),
	(i3) -- [scalar, ultra thick] (i4),
	(i5) -- [scalar, ultra thick] (i6),
	(i2) -- [fermion, ultra thick] (i4) -- [fermion, ultra thick] (i6) -- [fermion, ultra thick] (i8),

};
\end{feynman}
\node[red, thick] at (0.2,0.3) {$\varphi$};
\node[red, thick] at (2,0.3) {$\varphi$};
\node[red, thick] at (3.8,0.3) {$\varphi$};
\node[red, thick] at (0.2,-1) {$\Phi$};
\node[red, thick] at (3.8,-1) {$\Phi$};
\node[red, thick] at (0.2,-2.3) {$b$};
\node[red, thick] at (2,-2.3) {$b$};
\node[red, thick] at (3.8,-2.3) {$b$};
\end{tikzpicture}
\hspace{1cm}
\begin{tikzpicture}
\begin{feynman}
\vertex (i1);
\vertex [below=2cm of i1] (i5);
\vertex [right=1cm of i1] (i2);
\vertex [right=2cm of i2] (i3);
\vertex [right=1cm of i3] (i4);
\vertex [right=1cm of i5] (i6);
\vertex [right=2cm of i6] (i7);
\vertex [right=1cm of i7] (i8);
\diagram* {
	(i1) -- [scalar, ultra thick] (i2) -- [scalar, ultra thick] (i3) -- [scalar, ultra thick] (i4),
	(i2) -- [scalar, ultra thick] (i7),
	(i3) -- [scalar, ultra thick] (i6),
	(i5) -- [fermion, ultra thick] (i6) -- [fermion, ultra thick] (i7) -- [fermion, ultra thick] (i8),

};
\end{feynman}
\node[red, thick] at (0.2,0.3) {$\varphi$};
\node[red, thick] at (2,0.3) {$\varphi$};
\node[red, thick] at (3.8,0.3) {$\varphi$};
\node[red, thick] at (0.2,-1) {$\Phi$};
\node[red, thick] at (3.8,-1) {$\Phi$};
\node[red, thick] at (0.2,-2.3) {$b$};
\node[red, thick] at (2,-2.3) {$b$};
\node[red, thick] at (3.8,-2.3) {$b$};
\end{tikzpicture}
\hspace{1cm}
\begin{tikzpicture}
\begin{feynman}
\vertex (i1);
\vertex [below=2cm of i1] (i4);
\vertex [right=2cm of i1] (i2);
\vertex [right=2cm of i2] (i3);
\vertex [right=1cm of i4] (i5);
\vertex [right=2cm of i5] (i6);
\vertex [right=1cm of i6] (i7);

\diagram* {
	(i1) -- [scalar, ultra thick] (i2) -- [scalar, ultra thick] (i3),
	(i2) -- [scalar, ultra thick] (i5),
	(i2) -- [scalar, ultra thick] (i6),
	(i4) -- [fermion, ultra thick] (i5) -- [fermion, ultra thick] (i6) -- [fermion, ultra thick] (i7),

};
\end{feynman}
\node[red, thick] at (0.2,0.3) {$\varphi$};
\node[red, thick] at (3.8,0.3) {$\varphi$};
\node[red, thick] at (0.2,-1) {$\Phi$};
\node[red, thick] at (3.8,-1) {$\Phi$};
\node[red, thick] at (0.2,-2.3) {$b$};
\node[red, thick] at (2,-2.3) {$b$};
\node[red, thick] at (3.8,-2.3) {$b$};
\end{tikzpicture}
\caption{Box and triangle diagrams for the effective WIMP-bottom quark interactions with spin-0 mediator.}
\label{fig:scalar box}
\end{center}
\end{figure}

\subsubsection{Effective operator at the next-to-leading order}

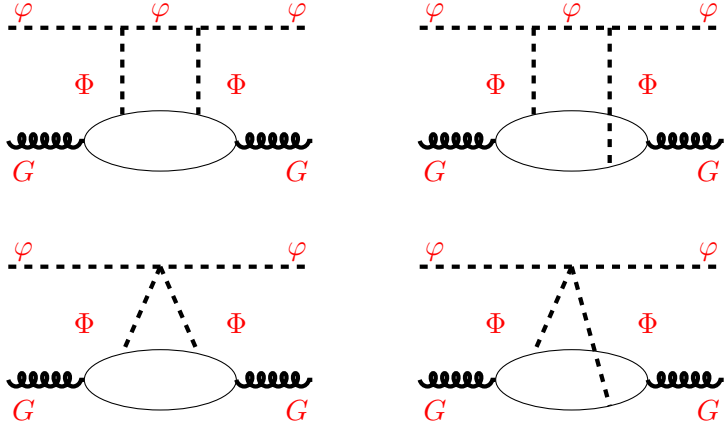
\begin{figure}[t]
\begin{center}
\begin{tikzpicture}
\begin{feynman}
\vertex (i1);
\vertex [right=1cm of i1] (i2);
\vertex [right=0.5cm of i2] (i3);
\vertex [right=1cm of i3] (i4);
\vertex [right=0.5cm of i4] (i5);
\vertex [right=1cm of i5] (i6);
\vertex [below=1.5cm of i1] (i7);
\vertex [below=1.5cm of i2] (i8);
\vertex [below=1.15cm of i3] (i9);
\vertex [below=1.15cm of i4] (i10);
\vertex [below=1.5cm of i5] (i11);
\vertex [below=1.5cm of i6] (i12);

\diagram* {
	(i1) -- [scalar, ultra thick] (i6),
	(i7) -- [gluon, ultra thick] (i8),
	(i11) -- [gluon, ultra thick] (i12),
	(i3) -- [scalar, ultra thick] (i9),
	(i4) -- [scalar, ultra thick] (i10),
};
\end{feynman}
\draw(2,-1.5) ellipse [x radius=1cm, y radius=0.4cm];
\node[red, thick] at (0.2,0.2) {$\varphi$};
\node[red, thick] at (2,0.2) {$\varphi$};
\node[red, thick] at (3.8,0.2) {$\varphi$};
\node[red, thick] at (1,-0.75) {$\Phi$};
\node[red, thick] at (3,-0.75) {$\Phi$};
\node[red, thick] at (0.2,-1.9) {$G$};
\node[red, thick] at (3.8,-1.9) {$G$};
\end{tikzpicture}
\hspace{1cm}
\begin{tikzpicture}
\begin{feynman}
\vertex (i1);
\vertex [right=1cm of i1] (i2);
\vertex [right=0.5cm of i2] (i3);
\vertex [right=1cm of i3] (i4);
\vertex [right=0.5cm of i4] (i5);
\vertex [right=1cm of i5] (i6);
\vertex [below=1.5cm of i1] (i7);
\vertex [below=1.5cm of i2] (i8);
\vertex [below=1.15cm of i3] (i9);
\vertex [below=1.845cm of i4] (i10);
\vertex [below=1.5cm of i5] (i11);
\vertex [below=1.5cm of i6] (i12);

\diagram* {
	(i1) -- [scalar, ultra thick] (i6),
	(i7) -- [gluon, ultra thick] (i8),
	(i11) -- [gluon, ultra thick] (i12),
	(i3) -- [scalar, ultra thick] (i9),
	(i4) -- [scalar, ultra thick] (i10),
};
\end{feynman}
\draw(2,-1.5) ellipse [x radius=1cm, y radius=0.4cm];
\node[red, thick] at (0.2,0.2) {$\varphi$};
\node[red, thick] at (2,0.2) {$\varphi$};
\node[red, thick] at (3.8,0.2) {$\varphi$};
\node[red, thick] at (1,-0.75) {$\Phi$};
\node[red, thick] at (3,-0.75) {$\Phi$};
\node[red, thick] at (0.2,-1.9) {$G$};
\node[red, thick] at (3.8,-1.9) {$G$};
\end{tikzpicture}
\\ \hspace*{\fill} \\
\begin{tikzpicture}
\begin{feynman}
\vertex (i1);
\vertex [right=1cm of i1] (i2);
\vertex [right=0.5cm of i2] (i3);
\vertex [right=0.5cm of i3] (i13);
\vertex [right=1cm of i3] (i4);
\vertex [right=0.5cm of i4] (i5);
\vertex [right=1cm of i5] (i6);
\vertex [below=1.5cm of i1] (i7);
\vertex [below=1.5cm of i2] (i8);
\vertex [below=1.15cm of i3] (i9);
\vertex [below=1.15cm of i4] (i10);
\vertex [below=1.5cm of i5] (i11);
\vertex [below=1.5cm of i6] (i12);

\diagram* {
	(i1) -- [scalar, ultra thick] (i6),
	(i7) -- [gluon, ultra thick] (i8),
	(i11) -- [gluon, ultra thick] (i12),
	(i13) -- [scalar, ultra thick] (i9),
	(i13) -- [scalar, ultra thick] (i10),
};
\end{feynman}
\draw(2,-1.5) ellipse [x radius=1cm, y radius=0.4cm];
\node[red, thick] at (0.2,0.2) {$\varphi$};
\node[red, thick] at (3.8,0.2) {$\varphi$};
\node[red, thick] at (1,-0.75) {$\Phi$};
\node[red, thick] at (3,-0.75) {$\Phi$};
\node[red, thick] at (0.2,-1.9) {$G$};
\node[red, thick] at (3.8,-1.9) {$G$};
\end{tikzpicture}
\hspace{1cm}
\begin{tikzpicture}
\begin{feynman}
\vertex (i1);
\vertex [right=1cm of i1] (i2);
\vertex [right=0.5cm of i2] (i3);
\vertex [right=0.5cm of i3] (i13);
\vertex [right=1cm of i3] (i4);
\vertex [right=0.5cm of i4] (i5);
\vertex [right=1cm of i5] (i6);
\vertex [below=1.5cm of i1] (i7);
\vertex [below=1.5cm of i2] (i8);
\vertex [below=1.15cm of i3] (i9);
\vertex [below=1.845cm of i4] (i10);
\vertex [below=1.5cm of i5] (i11);
\vertex [below=1.5cm of i6] (i12);

\diagram* {
	(i1) -- [scalar, ultra thick] (i6),
	(i7) -- [gluon, ultra thick] (i8),
	(i11) -- [gluon, ultra thick] (i12),
	(i13) -- [scalar, ultra thick] (i9),
	(i13) -- [scalar, ultra thick] (i10),
};
\end{feynman}
\draw(2,-1.5) ellipse [x radius=1cm, y radius=0.4cm];
\node[red, thick] at (0.2,0.2) {$\varphi$};
\node[red, thick] at (3.8,0.2) {$\varphi$};
\node[red, thick] at (1,-0.75) {$\Phi$};
\node[red, thick] at (3,-0.75) {$\Phi$};
\node[red, thick] at (0.2,-1.9) {$G$};
\node[red, thick] at (3.8,-1.9) {$G$};
\end{tikzpicture}
\end{center}
\caption{Two-loop diagrams for effective WIMP-gluon interactions with spin-0 mediator.}
\label{fig:scalar gluon}
\end{figure}

To derive the loop corrections to the spin-independent DM-nucleon scattering cross section, we first need to calculate the Wilson coefficients of  DM-quark and  DM-gluon  effective operators. Following Refs.~\cite{Hisano:2015bma,Hisano:2017jmz}, relevant effective operators take the form:
\begin{eqnarray}
 {\cal L}_{\rm eff} \sim C^{{\rm twist2}} \varphi^\dagger i\partial^\mu_{} i\partial^\nu_{}  \varphi {\cal O}_{\mu\nu}^b +  C^{{\rm scalar}} \varphi^\dagger \varphi m_b \bar b b + C^{\rm gluon} \varphi^\dagger \varphi {\alpha_s \over 12 \pi } G^a_{\mu\nu} G^{a\mu\nu} \label{masterx}
\end{eqnarray}
where ${\cal O}_{\mu\nu}^b$ is the twist-2 operator defined by 
\begin{eqnarray}
{\cal O}_{\mu\nu}^b = {i \over 2} \bar{b} (\partial_\mu \gamma_\nu + \partial_\nu\gamma_\mu - {1 \over 2}g_{\mu\nu}\slashed{\partial})b \; .
\end{eqnarray}
Wilson coefficients for scalar-type DM-bottom quark operator and the twist-2 operator, $C^{{\rm scalar}}$ and $C^{{\rm twist2}}$, are generated by the box and triangle diagrams in Fig.~\ref{fig:scalar box}. The two-loop diagrams in Fig.~\ref{fig:scalar gluon} with only bottom quark running in the loop contribute to the scalar-type DM-gluon operator with the Wilson coefficient $C^{{\rm gluon}}$.

Box and triangle diagrams are calculated in the zero-momentum transfer limit for simplification.  We further expand the integral by the bottom quark momentum and only keep the leading term~\cite{Abe:2018emu} in the calculation. Then the Wilson coefficients $C^{{\rm scalar}}$ and $C^{{\rm twist2}}$ can be obtained by reading out the DM-quark effective operators. For the DM-gluon coefficient $C^{{\rm gluon}}$, one needs to calculate the amplitude of two-loop diagrams and find the effective operator $\varphi^\dag \varphi G_{\mu\nu}^a G^{a\mu\nu}$. It should be noted that high twist DM-gluon effective operator, whose impacts to the DM-nuclei scattering cross section is sub-dominant, is neglected in this paper. We use the Fock-Schwinger gauge~\cite{Novikov:1983gd,Shtabovenko:2016sxi} for the gluon field, which makes the calculation transparent.  Then the two-loop amplitude can be expressed using the two-point function of scalar field in the gluon background field.  From this amplitude one can obtain the Wilson coefficient $C^{{\rm gluon}}$. All relevant Wilson coefficients are summarized as follows:
\begin{eqnarray}
C^{{\rm twist2}} &=& -{4 (f_p \Lambda )^2 \over 16\pi^2 } Z_{11}(m_\varphi^2, m_\varphi^2,m_\Phi^2)  \\
C^{{\rm scalar}} &=& {1 \over 16\pi^2 }\Big\{ (f_p \Lambda )^2\left[ -4 Z_{00}(m_\varphi^2,m_\varphi^2,m_\Phi^2) - m_{\varphi}^2 Z_{11}(m_\varphi^2,m_\varphi^2,m_\Phi^2)\right] \nonumber \\
&& + \lambda f_p^2 C_2(m_b^2,m_\Phi^2,m_b^2) \Big\} \\
C^{{\rm gluon}} &=& \frac{1}{32\pi^2}\left[ (f_P \Lambda)^2 \frac{\partial F_1(m_\Phi^2 )}{\partial m_\Phi^2} + \lambda f_P^2 \frac{\partial F_2(m_\Phi^2 )}{\partial m_\Phi^2} \right]
\end{eqnarray}
where the loop functions $Z_{11}$, $Z_{00}$, $F_1(m_\Phi^2)$ and $F_2(m_\Phi^2)$ are given in the Appendix \ref{loop}. In our numerical calculations, we use Package-X \cite{Passarino:1978jh, Abe:2015rja} to compute the above loop functions.

\subsubsection{Effects from RGE running and Threshold Matching}
The effective operators and Wilson coefficients obtained above are defined at the electroweak scale $\mu=M_{Z}$ and need to evolve to the hadron scale $\mu_{\rm hadron} \approx 1~{\rm GeV}$.  They evolve by means of renormalization group equations (RGEs). As  shown in the Eq.~(\ref{masterx}), one needs to account the running of scalar-type operators and a twist-2 operator.   The QED contributions to the renormalization group evolution can be neglected taking into account the smallness of the electromagnetic coupling constant.  Notice that DM is a QCD and QED singlet, so the RGE evolution is mainly due to the SM fields in the effective operators.  The quark mass operator has vanishing anomalous dimension in a mass-independent renormalization scheme like the $\overline{\rm MS}$ scheme~\cite{Hisano:2015bma}, 
\begin{eqnarray}
\mu { d \over d \mu } m_q \bar q q =0 \; .
\end{eqnarray}
By differentiating the trace of energy-momentum tensor in QCD, one further has $\mu {d\over d \mu}\left( {\alpha_s \over \pi }GG\right)$ $ =0$, which implies the Wilson coefficients of scalar-type operators are RGE invariant. 
Finite corrections arise at the bottom quark threshold, $\mu=\mu_b$,
\begin{eqnarray}
C^{\rm gluon} |_{\mu=\mu_b} \to C^{\rm gluon} |_{\mu=\mu_b}-C^{{\rm scalar}}|_{\mu=\mu_b}\; ,
\end{eqnarray}
and also there is threshold corrections to $\alpha_s$~\cite{Chetyrkin:1997un}. As a result, the effect of the DM-bottom-quark operator appears as  an extra contribution to the DM-gluon effective operator.

To evaluate the RGE effect of  the twist-2 effective operator in Eq.~(\ref{masterx}), we need to introduce the twist-2 operator of gluon with the Wilson $C_g^{\rm twist2}$. The relevant RGE is~\cite{Gross:1974cs}
\begin{eqnarray}
{d \over d \ln \mu}\left( C_q^{\rm twist2},~C_{g}^{\rm twist2} \right) = \left( C_q^{\rm twist2},~C_{g}^{\rm twist2} \right)  \Gamma \label{rgerun}
\end{eqnarray}  
with
\begin{eqnarray}
\Gamma= {\alpha_s \over 4 \pi} \begin{pmatrix}  {16\over 3} C_F & 0 & \cdots &0 & {4\over 3} \cr  0& {16\over 3} C_F & 0 & \cdots & {4\over 3} \cr \cdots & \cdots & \cdots & \cdots &\cdots \cr  {16\over 3} C_F & \cdots & \cdots & {16\over 3} C_F  & {4 \over 3} N_f\end{pmatrix}
\end{eqnarray}
where $C_F=4/3$ being the quadratic Casimir invariant and $N_f$ is the number of quark family.  $\Gamma$ is a $(N_f+1)\times(N_f+1)$ matrix. Wilson coefficients of twist-2 operators cross thresholds continuously.

\subsubsection{Scattering cross section}
The spin-independent scattering cross section can then be written as 
\begin{eqnarray}
\sigma^{\rm SI}={\mu^2 \over 8\pi } {m_N^2 \over m_\varphi^2} \left[ \left(C^{\rm scalar} -C^{\rm gluon} \right) f_{T_b}^{N}  +\sum_q  C_q^{\rm twist2} {3\over 4} m_\varphi^2 (\bar q^N+ q^N) -{3\over 4} C_g^{\rm twist2}  m_\varphi^2 G^N  \right]^2,  \nonumber  \\ \label{masterr}
\end{eqnarray}
where
\begin{eqnarray}
f_{T_b}^N= {2\over 27} \left(1-\sum_{u,d,s} f_{T_q}^N \right)={2\over 27 } f_{T_G}^N \; ,
\end{eqnarray}
is the nucleon matrix element of bottom quark, $q^N$, $\bar q^N$ and $G^N$ is the second moment of quarks, antiquarks and gluon in the nucleon, $m_N$ is the nucleon mass and $\mu$ is the reduced mass of the WIMP-nucleon system. The numerical values of nucleon form factors are given in the Appendix~\ref{nucleon matrix elements}. Non-zero twist-2 Wilson coefficients for light quarks and gluon come from the RGE running  in the Eq.~(\ref{rgerun}).

\section{Real scalar DM with spin-1 mediator}

\begin{figure}[t]
\begin{center}
\begin{tikzpicture}
\begin{feynman}
\vertex (i1);
\vertex [right=0.75cm of i1] (i2);
\vertex [right=1.5cm of i2] (i3);
\vertex [right=0.75cm of i3] (i4);
\vertex [below=1.5cm of i1] (i5);
\vertex [below=0.75cm of i2] (i6);
\vertex [below=0.75cm of i3] (i7);
\vertex [below=1.5cm of i4] (i8);
\diagram* {
	(i1) -- [scalar, ultra thick] (i6) -- [scalar, ultra thick] (i5),
	(i6) -- [photon, ultra thick] (i7),
	(i8) -- [fermion, ultra thick] (i7) -- [fermion, ultra thick] (i4),
};
\end{feynman}
\node[red, thick] at (0.2,0.3) {$\rho,\eta$};
\node[red, thick] at (0.2,-1.8) {$\rho,\eta$};
\node[red, thick] at (1.5,-0.4) {$Z'$};
\node[red, thick] at (2.8,0.3) {$b$};
\node[red, thick] at (2.8,-1.8) {$\bar{b}$};
\end{tikzpicture}
\hspace{0.75cm}
\begin{tikzpicture}
\begin{feynman}
\vertex (i1);
\vertex [below=1.5cm of i1] (i4);
\vertex [right=1.5cm of i1] (i2);
\vertex [right=1.5cm of i2] (i3);
\vertex [right=1.5cm of i4] (i5);
\vertex [right=1.5cm of i5] (i6);

\diagram* {
	(i1) -- [scalar, ultra thick] (i2) -- [photon, ultra thick] (i3),
	(i4) -- [scalar, ultra thick] (i5) -- [photon, ultra thick] (i6),
	(i2) -- [scalar, ultra thick] (i5)

};
\end{feynman}
\node[red, thick] at (0.2,0.3) {$\rho,\eta$};
\node[red, thick] at (0.2,-1.8) {$\rho,\eta$};
\node[red, thick] at (1.0,-0.75) {$\eta,\rho$};
\node[red, thick] at (2.8,0.3) {$Z'$};
\node[red, thick] at (2.8,-1.8) {$Z'$};
\end{tikzpicture}
\hspace{0.75cm}
\begin{tikzpicture}
\begin{feynman}
\vertex (i1);
\vertex [below=1.5cm of i1] (i4);
\vertex [right=1.5cm of i1] (i2);
\vertex [right=1.5cm of i2] (i3);
\vertex [right=1.5cm of i4] (i5);
\vertex [right=1.5cm of i5] (i6);

\diagram* {
	(i1) -- [scalar, ultra thick] (i2) -- [photon, ultra thick] (i6),
	(i4) -- [scalar, ultra thick] (i5) -- [photon, ultra thick] (i3),
	(i2) -- [scalar, ultra thick] (i5)
};
\end{feynman}
\node[red, thick] at (0.2,0.3) {$\rho,\eta$};
\node[red, thick] at (0.2,-1.8) {$\rho,\eta$};
\node[red, thick] at (1.0,-0.75) {$\eta,\rho$};
\node[red, thick] at (2.8,0.3) {$Z'$};
\node[red, thick] at (2.8,-1.8) {$Z'$};
\end{tikzpicture}
\hspace{0.75cm}
\begin{tikzpicture}
\begin{feynman}
\vertex (i1);
\vertex [right=0.75cm of i1] (i2);
\vertex [right=0.75cm of i2] (i3);
\vertex [below=1.5cm of i1] (i4);
\vertex [below=0.75cm of i2] (i5);
\vertex [below=1.5cm of i3] (i6);

\diagram* {
	(i1) -- [scalar, ultra thick] (i5) -- [photon, ultra thick] (i3),
	(i4) -- [scalar, ultra thick] (i5) -- [photon, ultra thick] (i6),
};
\end{feynman}
\node[red, thick] at (0,0.3) {$\rho,\eta$};
\node[red, thick] at (0,-1.8) {$\rho,\eta$};
\node[red, thick] at (1.5,0.3) {$Z'$};
\node[red, thick] at (1.5,-1.8) {$Z'$};
\end{tikzpicture}
\end{center}
\caption{Annihilation channels of real scalar dark matter $\rho$ with a spin-1 mediator $Z'$.}
\label{fig:vector annihilation}
\end{figure}
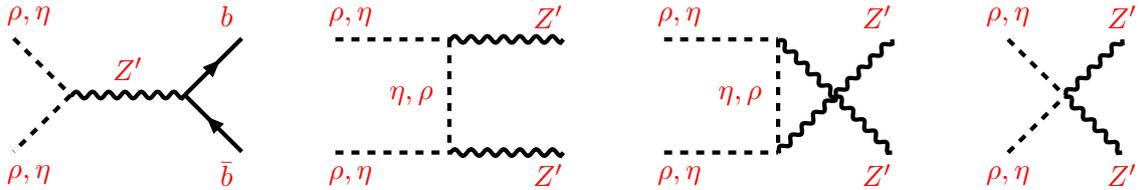

In this section we assume a complex scalar  $\varphi$ couples to the SM via a spin-1 mediator $Z^\prime_\mu$, which can be either a new gauge boson for a spontaneously broken $U(1)$ gauge symmetry  or a matter field. Here we take $Z^\prime$ as a gauge boson for simplicity. The Lagrangian for the $\varphi$ can be written as 
\begin{eqnarray}
{\cal L} \sim (D_\mu \varphi)^\dagger  (D^\mu \varphi ) + (D_\mu \phi)^\dagger  (D^\mu \phi ) -V(\varphi, \phi) - \zeta\bar b \gamma^\mu_{} b Z^\prime_\mu  \label{vectorx}
\end{eqnarray}
where $\phi$ is the scalar boson whose vacuum expectation value leads to the spontaneous breaking of the $U(1)$ gauge symmetry as well as the origin of  the $Z^\prime$ mass, $\zeta$ equals to the new gauge coupling $g_V$ times the hypercharge of the bottom quark, and
\begin{eqnarray}
D_\mu=\partial_\mu-ig_V^{} Z^\prime_\mu \; .
\end{eqnarray}
The scalar potential can be written as
\begin{eqnarray}
V(\phi, \varphi) &=& -\mu_\phi^2 \phi^\dagger \phi   + \mu_\varphi^2 \varphi^\dagger \varphi+ \lambda_\phi (\phi^\dagger \phi)^2 + \lambda_\varphi (\varphi^\dagger \varphi)^2  \nonumber \\
&&+ \lambda_{\varphi \phi} \varphi^\dagger \varphi \phi^\dagger \phi + \lambda_{\varphi \phi}^\prime ( \varphi^2 + h.c.) \phi^\dagger \phi \; ,
\end{eqnarray}
where $\varphi=(\rho+ i\eta)/\sqrt{2}$ and $\phi=(v_\phi + \sigma+ i G)/\sqrt{2}$. The potential has a $Z_2$ discrete symmetry,  $\varphi\leftrightarrow -\varphi$, which stabilizes $\varphi$.
Due to the last term in the potential, there might be mass splitting between the CP-even and the CP-odd components of $\varphi$,  
\begin{eqnarray}
m_{\rho,\eta}^2 = \mu_\varphi^2 + {1\over 2 } \lambda_{\varphi \phi} v_\phi^2 \pm \lambda_{\varphi \phi}^\prime v_\phi^2
\end{eqnarray}
and only the lighter component (here we take it as the CP-even component $\rho$) can serve as the dark matter candidate, which is quite similar to the case of the inert dark matter in two-Higgs doublet models~\cite{LopezHonorez:2006gr}.  $\eta$ will decay into $\rho$ plus SM particles.  We further assume that $\sigma$ is much heavier than $\rho$, since the case of scalar portal has already been studied in the last section and we will not study the $\sigma$ portal in this case. As a result, $\rho$ mainly annihilate into $\bar b b$ and $Z^\prime Z^\prime$ final states, which will be discussed in the following subsection.  It should be noted that the SM Higgs is supposed to be decoupled from the hidden Higgs sector for simplicity.  A systematic study of  the scalar mass spectrum and various constraints from precision measurements  for a general Higgs potential is beyond the reach of this paper. We refer the reader to Ref.~\cite{Barger:2007im}  and references cited therein for details.

As can be seen from the Eq.~(\ref{vectorx}),  $Z^\prime$ couples to the vector-bilinear of bottom quark.  Such an interaction is true in the $U(1)_{B-L}$~\cite{Mohapatra:1980qe}, $U(1)_{B+L}$~\cite{Chao:2016avy} and $U(1)_{B_i-B_3}$~\cite{He:1991qd} models. Since the main purpose of this paper is to estimate the loop corrections to the direct detection cross section, we will not focus on any specific model. Corrections induced by axial-vector current is similar to the vector current case.

\subsection{Relic density}
In this subsection we calculate the relic density of  scalar DM in the vector-portal.  As mentioned above, we are interested in the scenario where $\eta$ is slightly heavier than $\rho$. As a result, one needs to include the co-annihilation~\cite{Edsjo:1997bg} effect and the DM number density can be written as, \begin{eqnarray}
n_{\rm DM} = n_\rho+n_\eta \; ,
\end{eqnarray}
where $n_\rho$ and $n_\eta$ are number densities of $\rho$ and $\eta$ respectively.  
 
\subsubsection{The  $m_{\rho, \eta}<m_{Z'}$ scenario}
In the mass regime $ m_{\rho,\eta}<m_{Z'}$, the annihilation channel $\rho \rho( \eta \eta)\to Z'Z'$ is kinematically forbidden and there is only one annihilation channel $\rho \eta \to\bar b b$ which is given in  the most left-panel of the Fig. \ref{fig:vector annihilation}. The annihilation cross section can be written as 
\begin{equation}
\begin{split}
\sigma \approx \frac{\zeta^2g_V^2}{12\pi s}\frac{(s-4m_{\rho}^2)(s+2m_b^2)}{ (s-m_{Z'}^2)^2+m_{Z'}^2\Gamma_{Z'}^2}\sqrt{\frac{s-4m_b^2}{s-4m_\rho^2}}
\end{split}
\end{equation}
where we have neglected the correction induced by the mass difference, and $\Gamma_{Z'}$  is the decay rate of $Z'$, 
\begin{equation}
\begin{split}
\Gamma_{Z'} = & \frac{\zeta^2}{4\pi}\left(1+\frac{2m_b^2}{m_{Z'}^2}\right)\sqrt{m_{Z'}^2-4m_b^2} \\
              & + \Theta(m_{Z'}-2m_\rho) \frac{g_V^2}{16\pi}\left(1-\frac{4m_\rho^2}{m_{Z'}^2}\right)\sqrt{m_{Z'}^2-4m_\rho^2}
\end{split} 
\end{equation}
with $\Theta(x)$ the unit step function. Analytically one can approximate the thermal average $\langle \sigma v\rangle$ with the non-relativistic expansion $\langle \sigma v \rangle =a+b\langle v^2 \rangle$ in the laboratory frame, where $a$ and $b$ are given in the Appendix \ref{annihilation}.

\subsubsection{The  $m_{\rho,\eta}>m_{Z'}$ scenario}

In the mass regime $ m_{\rho,\eta}>m_{Z'}$, the annihilation channel $\rho \rho (\eta\eta) \to Z'Z'$, as shown in the remaining panels of Fig. \ref{fig:vector annihilation}, is kinematically allowed. The annihilation cross section in the center-of-mass framework  is 
\begin{equation}
\begin{split}
\sigma = \frac{g_V^4}{16\pi s} \Bigg[ 4 + \frac{(m_{Z'}^2 - 4m_{\varphi}^2)^2}{A^2 - B^2} + \frac{4s(m_{Z'}^2-2m_\varphi^2)+(m_{Z'}^2-4m_\varphi^2)^2}{AB}\tanh^{-1}\left(\frac{B}{A}\right) \Bigg]
\end{split}
\end{equation}
where $A=s/2-m_{Z'}^2$ and $B=1/2\sqrt{(s-4m_\varphi^2)(s-4m_{Z'}^2)} $ with $\varphi=\rho,\eta$. The expressions of $s$-wave and $p$-wave contributions are given in the Appendix \ref{annihilation}, with the help of which one may estimate the relic abundance of DM.

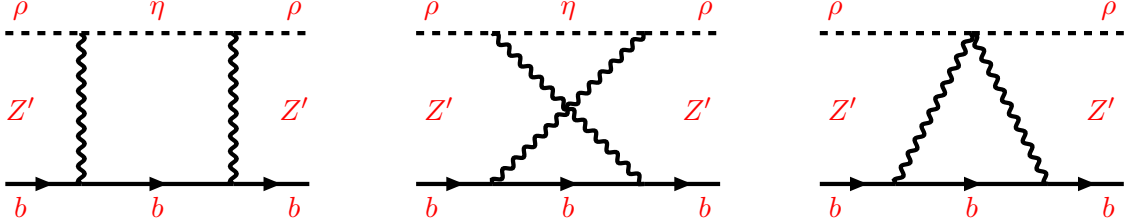
\begin{figure}[htbp]
\begin{center}
\begin{tikzpicture}
\begin{feynman}
\vertex (i1);
\vertex [below=2cm of i1] (i2);
\vertex [right=1cm of i1] (i3);
\vertex [right=1cm of i2] (i4);
\vertex [right=2cm of i3] (i5);
\vertex [right=2cm of i4] (i6);
\vertex [right=1cm of i5] (i7);
\vertex [right=1cm of i6] (i8);
\diagram* {
	(i1) -- [scalar,ultra thick] (i3) -- [scalar, ultra thick] (i5) -- [scalar, ultra thick] (i7),
	(i3) -- [photon, ultra thick] (i4),
	(i5) -- [photon, ultra thick] (i6),
	(i2) -- [fermion, ultra thick] (i4) -- [fermion, ultra thick] (i6) -- [fermion, ultra thick] (i8),

};
\end{feynman}
\node[red, thick] at (0.2,0.3) {$\rho$};
\node[red, thick] at (2,0.3) {$\eta$};
\node[red, thick] at (3.8,0.3) {$\rho$};
\node[red, thick] at (0.2,-1) {$Z'$};
\node[red, thick] at (3.8,-1) {$Z'$};
\node[red, thick] at (0.2,-2.3) {$b$};
\node[red, thick] at (2,-2.3) {$b$};
\node[red, thick] at (3.8,-2.3) {$b$};
\end{tikzpicture}
\hspace{1cm}
\begin{tikzpicture}
\begin{feynman}
\vertex (i1);
\vertex [below=2cm of i1] (i5);
\vertex [right=1cm of i1] (i2);
\vertex [right=2cm of i2] (i3);
\vertex [right=1cm of i3] (i4);
\vertex [right=1cm of i5] (i6);
\vertex [right=2cm of i6] (i7);
\vertex [right=1cm of i7] (i8);
\diagram* {
	(i1) -- [scalar, ultra thick] (i2) -- [scalar, ultra thick] (i3) -- [scalar, ultra thick] (i4),
	(i2) -- [photon, ultra thick] (i7),
	(i3) -- [photon, ultra thick] (i6),
	(i5) -- [fermion, ultra thick] (i6) -- [fermion, ultra thick] (i7) -- [fermion, ultra thick] (i8),

};
\end{feynman}
\node[red, thick] at (0.2,0.3) {$\rho$};
\node[red, thick] at (2,0.3) {$\eta$};
\node[red, thick] at (3.8,0.3) {$\rho$};
\node[red, thick] at (0.3,-1) {$Z'$};
\node[red, thick] at (3.7,-1) {$Z'$};
\node[red, thick] at (0.2,-2.3) {$b$};
\node[red, thick] at (2,-2.3) {$b$};
\node[red, thick] at (3.8,-2.3) {$b$};
\end{tikzpicture}
\hspace{1cm}
\begin{tikzpicture}
\begin{feynman}
\vertex (i1);
\vertex [below=2cm of i1] (i4);
\vertex [right=2cm of i1] (i2);
\vertex [right=2cm of i2] (i3);
\vertex [right=1cm of i4] (i5);
\vertex [right=2cm of i5] (i6);
\vertex [right=1cm of i6] (i7);

\diagram* {
	(i1) -- [scalar, ultra thick] (i2) -- [scalar, ultra thick] (i3),
	(i2) -- [photon, ultra thick] (i5),
	(i2) -- [photon, ultra thick] (i6),
	(i4) -- [fermion, ultra thick] (i5) -- [fermion, ultra thick] (i6) -- [fermion, ultra thick] (i7),

};
\end{feynman}
\node[red, thick] at (0.2,0.3) {$\rho$};
\node[red, thick] at (3.8,0.3) {$\rho$};
\node[red, thick] at (0.3,-1) {$Z'$};
\node[red, thick] at (3.7,-1) {$Z'$};
\node[red, thick] at (0.2,-2.3) {$b$};
\node[red, thick] at (2,-2.3) {$b$};
\node[red, thick] at (3.8,-2.3) {$b$};
\end{tikzpicture}
\caption{Box and triangle diagrams for the effective WIMP-bottom quark interactions with spin-1 mediator.}
\label{fig:vector box}
\end{center}
\end{figure}

\subsection{Direct detection}
Since DM is non-relativistic, the  process $ \rho + N \to \eta+N$ is kinematically forbidden at the tree level.  Although the process $\eta+N\to N+\rho$ is allowed, the number density of $\eta$ is negligibly small, which leads to null signal in direct detection experiments unless there is a mechanism that can produce $\eta$ near the Earth. DM-nucleon scatterings arise at the loop level in this scenario. 
\subsubsection{Effective operators at loop level}
\begin{figure}
\begin{center}
\begin{tikzpicture}
\begin{feynman}
\vertex (i1);;
\vertex [right=1cm of i1] (i2);
\vertex [right=0.5cm of i2] (i3);
\vertex [right=1cm of i3] (i4);
\vertex [right=0.5cm of i4] (i5);
\vertex [right=1cm of i5] (i6);
\vertex [below=1.5cm of i1] (i7);
\vertex [below=1.5cm of i2] (i8);
\vertex [below=1.15cm of i3] (i9);
\vertex [below=1.15cm of i4] (i10);
\vertex [below=1.5cm of i5] (i11);
\vertex [below=1.5cm of i6] (i12);

\diagram* {
	(i1) -- [scalar, ultra thick] (i6),
	(i7) -- [gluon, ultra thick] (i8),
	(i11) -- [gluon, ultra thick] (i12),
	(i3) -- [photon, ultra thick] (i9),
	(i4) -- [photon, ultra thick] (i10),
};
\end{feynman}
\draw(2,-1.5) ellipse [x radius=1cm, y radius=0.4cm];
\node[red, thick] at (0.2,0.2) {$\rho$};
\node[red, thick] at (3.8,0.2) {$\rho$};
\node[red, thick] at (2,0.2) {$\eta$};
\node[red, thick] at (1,-0.75) {$Z'$};
\node[red, thick] at (3,-0.75) {$Z'$};
\node[red, thick] at (0.2,-1.9) {$G$};
\node[red, thick] at (3.8,-1.9) {$G$};
\end{tikzpicture}
\hspace{1cm}
\begin{tikzpicture}
\begin{feynman}
\vertex (i1);;
\vertex [right=1cm of i1] (i2);
\vertex [right=0.5cm of i2] (i3);
\vertex [right=1cm of i3] (i4);
\vertex [right=0.5cm of i4] (i5);
\vertex [right=1cm of i5] (i6);
\vertex [below=1.5cm of i1] (i7);
\vertex [below=1.5cm of i2] (i8);
\vertex [below=1.15cm of i3] (i9);
\vertex [below=1.845cm of i4] (i10);
\vertex [below=1.5cm of i5] (i11);
\vertex [below=1.5cm of i6] (i12);

\diagram* {
	(i1) -- [scalar, ultra thick] (i6),
	(i7) -- [gluon, ultra thick] (i8),
	(i11) -- [gluon, ultra thick] (i12),
	(i3) -- [photon, ultra thick] (i9),
	(i4) -- [photon, ultra thick] (i10),
};
\end{feynman}
\draw(2,-1.5) ellipse [x radius=1cm, y radius=0.4cm];
\node[red, thick] at (0.2,0.2) {$\rho$};
\node[red, thick] at (3.8,0.2) {$\rho$};
\node[red, thick] at (2,0.2) {$\eta$};
\node[red, thick] at (1,-0.75) {$Z'$};
\node[red, thick] at (3,-0.75) {$Z'$};
\node[red, thick] at (0.2,-1.9) {$G$};
\node[red, thick] at (3.8,-1.9) {$G$};
\end{tikzpicture}
\\ \hspace*{\fill} \\
\begin{tikzpicture}
\begin{feynman}
\vertex (i1);;
\vertex [right=1cm of i1] (i2);
\vertex [right=0.5cm of i2] (i3);
\vertex [right=0.5cm of i3] (i13);
\vertex [right=1cm of i3] (i4);
\vertex [right=0.5cm of i4] (i5);
\vertex [right=1cm of i5] (i6);
\vertex [below=1.5cm of i1] (i7);
\vertex [below=1.5cm of i2] (i8);
\vertex [below=1.15cm of i3] (i9);
\vertex [below=1.15cm of i4] (i10);
\vertex [below=1.5cm of i5] (i11);
\vertex [below=1.5cm of i6] (i12);

\diagram* {
	(i1) -- [scalar, ultra thick] (i6),
	(i7) -- [gluon, ultra thick] (i8),
	(i11) -- [gluon, ultra thick] (i12),
	(i13) -- [photon, ultra thick] (i9),
	(i13) -- [photon, ultra thick] (i10),
};
\end{feynman}
\draw(2,-1.5) ellipse [x radius=1cm, y radius=0.4cm];
\node[red, thick] at (0.2,0.2) {$\rho$};
\node[red, thick] at (3.8,0.2) {$\rho$};
\node[red, thick] at (1,-0.75) {$Z'$};
\node[red, thick] at (3,-0.75) {$Z'$};
\node[red, thick] at (0.2,-1.9) {$G$};
\node[red, thick] at (3.8,-1.9) {$G$};
\end{tikzpicture}
\hspace{1cm}
\begin{tikzpicture}
\begin{feynman}
\vertex (i1);;
\vertex [right=1cm of i1] (i2);
\vertex [right=0.5cm of i2] (i3);
\vertex [right=0.5cm of i3] (i13);
\vertex [right=1cm of i3] (i4);
\vertex [right=0.5cm of i4] (i5);
\vertex [right=1cm of i5] (i6);
\vertex [below=1.5cm of i1] (i7);
\vertex [below=1.5cm of i2] (i8);
\vertex [below=1.15cm of i3] (i9);
\vertex [below=1.845cm of i4] (i10);
\vertex [below=1.5cm of i5] (i11);
\vertex [below=1.5cm of i6] (i12);

\diagram* {
	(i1) -- [scalar, ultra thick] (i6),
	(i7) -- [gluon, ultra thick] (i8),
	(i11) -- [gluon, ultra thick] (i12),
	(i13) -- [photon, ultra thick] (i9),
	(i13) -- [photon, ultra thick] (i10),
};
\end{feynman}
\draw(2,-1.5) ellipse [x radius=1cm, y radius=0.4cm];
\node[red, thick] at (0.2,0.2) {$\rho$};
\node[red, thick] at (3.8,0.2) {$\rho$};
\node[red, thick] at (1,-0.75) {$Z'$};
\node[red, thick] at (3,-0.75) {$Z'$};
\node[red, thick] at (0.2,-1.9) {$G$};
\node[red, thick] at (3.8,-1.9) {$G$};
\end{tikzpicture}
\end{center}
\caption{Two-loop diagrams for effective WIMP-gluon interactions for spin-1 mediator.}
\label{fig:vector gluon}
\end{figure}
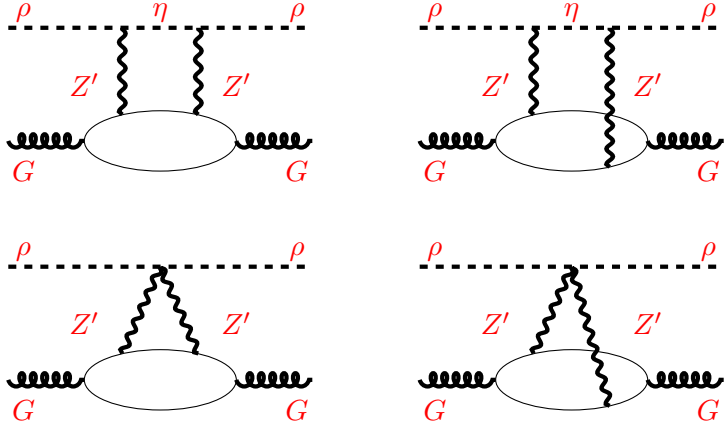

The effective DM-quark interactions can be generated at loop-level by box diagrams or the triangle diagram, as shown in the Fig.~\ref{fig:vector box}.  The DM-gluon interaction is generated at the two-loop level in the Fig.~\ref{fig:vector gluon}.
The effective Lagrangian can be written as
\begin{eqnarray}
 {\cal L}_{\rm eff} \sim C^{{\rm twist2}} \rho i\partial^\mu_{} i\partial^\nu_{}  \rho {\cal O}_{\mu\nu}^b +  C^{{\rm scalar}} \rho^2 m_b \bar b b + C^{\rm gluon} \rho^2 {\alpha_s \over 12 \pi } G^a_{\mu\nu} G^{a\mu\nu} \; .
\end{eqnarray}
Wilson coefficients for effective  DM-quark operators are given as 
\begin{eqnarray}
C^{\rm twist2} &=& {(\zeta g_V)^2 \over 16\pi^2}\Big[16X_1(m_\varphi^2, m_\varphi^2, 0, m_{Z'}^2)-16m_{\varphi}^2Z_{11}(m_\varphi^2, m_\varphi^2, m_{Z'}^2) \nonumber \\
               & & -32Z_{00}(m_\varphi^2, m_\varphi^2, m_{Z'}^2)\Big] \\
C^{\rm scalar} &=& {(\zeta g_V)^2 \over 16\pi^2 } \Big\{{1 \over 4} m_\varphi^2 \Big[ 16X_1(m_\varphi^2, m_\varphi^2, 0, m_{Z'}^2)-16m_{\varphi}^2Z_{11}(m_\varphi^2, m_\varphi^2, m_{Z'}^2) \nonumber \\
               & & -32Z_{00}(m_\varphi^2, m_\varphi^2, m_{Z'}^2)\Big] + \Big[ 16m_{\varphi}^2Z_{00}(m_\varphi^2, m_\varphi^2, m_{Z'}^2) + 2C_2(m_b^2, m_{Z'}^2, m_b^2) \nonumber \\
               & & - 2C_0(m_b^2, m_{Z'}^2, m_b^2) \Big] \Big\} 
\end{eqnarray}
To calculate the Wilson coefficient for the DM-gluon interaction, one needs  the two-point function of $Z^\prime$ in the gluon background field, which can be written as
\begin{eqnarray}
i \Pi_{Z^\prime Z^\prime}^{\alpha \beta}(q^2) = - {i \zeta^2\alpha_s \over 12\pi} G_{\mu\nu}^a G^{a\mu\nu} \left(- {g^{\alpha \beta }\over q^2} + {q^\alpha q^\beta \over q^4} \right)
\end{eqnarray}
Then the Wilson coefficient of DM-gluon operator is
\begin{eqnarray}
C^{\rm gluon} &=& {(\zeta g_V)^2 \over 16\pi^2} \Big\{ 4m_\varphi^2 \Big[ -X_2(m_\varphi^2, m_\varphi^2, 0, m_{Z'}^2) + Z_{00}(m_\varphi^2, m_\varphi^2, m_{Z'}^2) \nonumber \\
& & + m_\varphi^2 Z_{11}(m_\varphi^2, m_\varphi^2, m_{Z'}^2)  \Big] + 6C_0(0, m_{Z'}^2, 0) \Big\} \; .
\end{eqnarray}

Note that the RGE running of Wilson coefficients are the same as those discussed in the last section. The DM-nucleon cross section is the same as Eq.~(\ref{masterr}) up to replacements of Wilson coefficients.

\section{Numerical results}
\begin{figure}[htbp]
\centering
\includegraphics[scale=0.75]{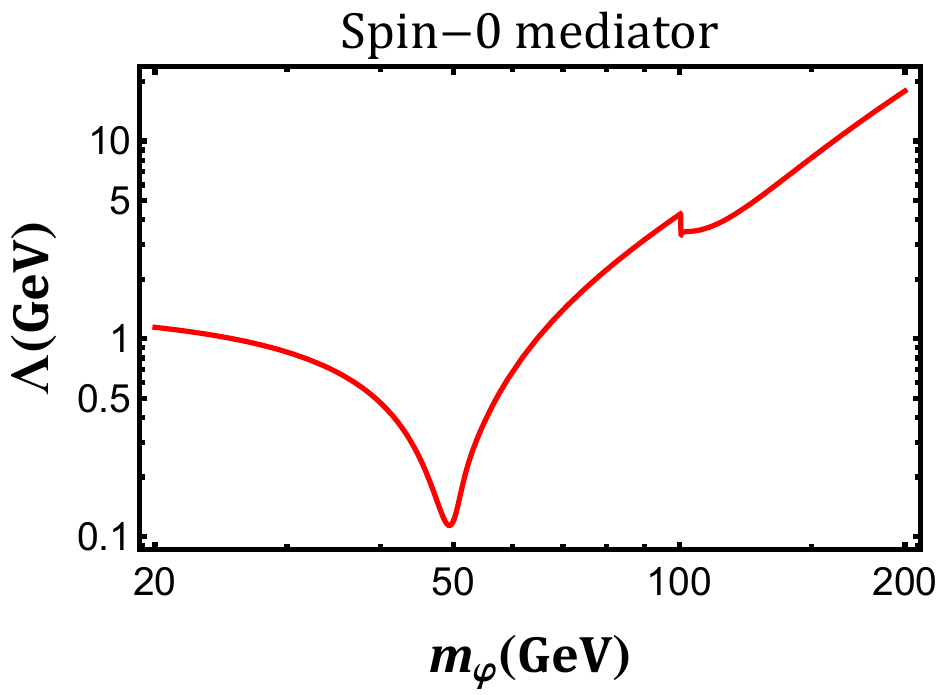}
\hspace{0.5cm}
\includegraphics[scale=0.75]{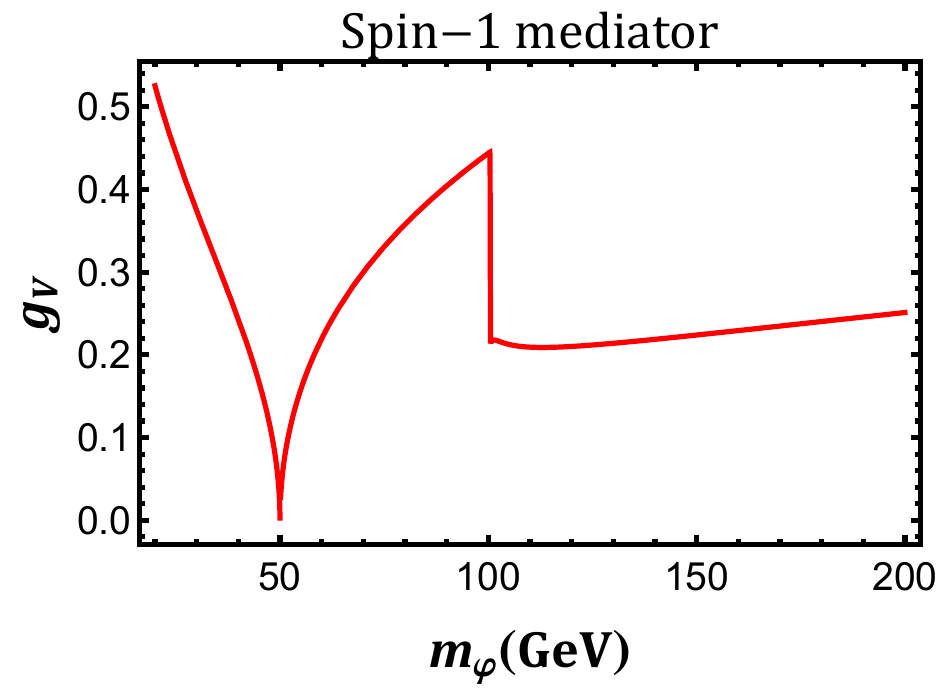}
\caption{Left panel: Coupling $\Lambda$ as the function of the DM mass $m_{\varphi}$ by setting $f_{P} = 1.5$, $\lambda = 0.1$ and $m_{\rm \Phi} = 100\,{\rm GeV}$. Right panel: New gauge coupling $g_{V}$ as the function of $m_\varphi$ by setting $\zeta = g_{V}$ and $m_{Z'} = 100\,{\rm GeV}$. Both are constrained by the observed dark matter relic density $\Omega h^2 = 0.1198$.}
\label{fig:coupling}
\end{figure}

In this section we present the numerical results for our models. We start by determining the couplings using the measured DM relic density $\Omega h^2 = 0.1198$. For the spin-0 mediator case, we show in the left panel of the Fig.~\ref{fig:coupling} the coupling $\Lambda$ as the function of  dark matter mass $m_{\varphi}$, by setting $f_P = 1.5$, $\lambda = 0.1$ and $m_{\Phi} = 100\,{\rm GeV}$. The first dip appears at  $m_{\varphi} \sim m_{\Phi}/2$, where the annihilation $\varphi^\dag \varphi \rightarrow \bar{b}b$ is resonantly enhanced. The second dip at $m_{\varphi} \sim m_{\Phi}$ is due to the opening of the annihilation channel $\varphi^\dag\varphi \rightarrow \Phi\Phi$. The similar procedure can be applied to the spin-1 mediator case, for which the new gauge coupling $g_{V}$ as the function of dark matter mass is shown in the right panel of the Fig.~\ref{fig:coupling}, by setting $\zeta = g_{V}$ and $m_{Z'} = 100\,{\rm GeV}$. It should be noted that there is only one viable annihilation channel $\varphi^{\dag}\varphi \rightarrow \bar{b}b$ for $m_{\varphi} < m_{Z'}$ which is $p$-wave suppressed, while $\varphi^{\dag}\varphi \rightarrow Z'Z'$ has both $s$-wave and $p$-wave components, so the second dip at $m_{\varphi}=m_{Z'}$ is significant.

\begin{figure}[htbp]
\centering
\includegraphics[scale=0.72]{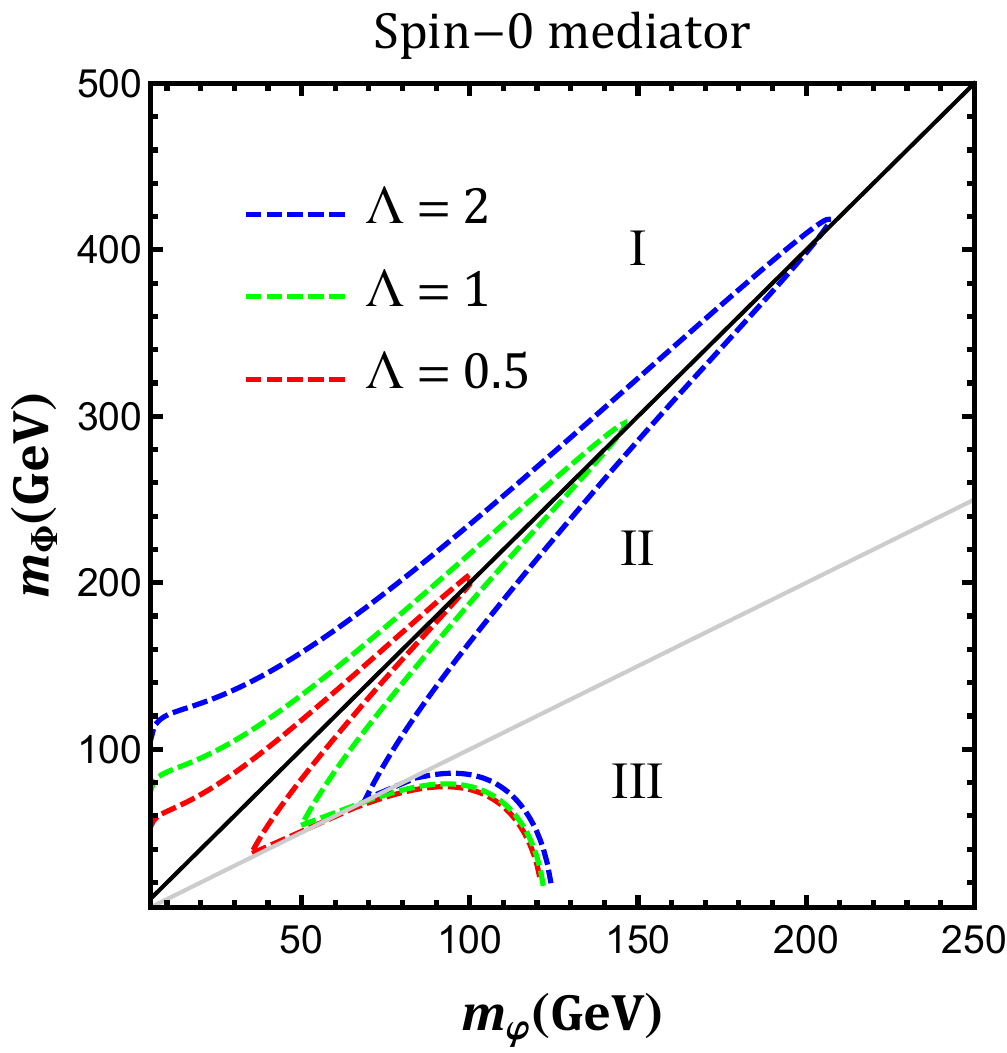}
\includegraphics[scale=0.87]{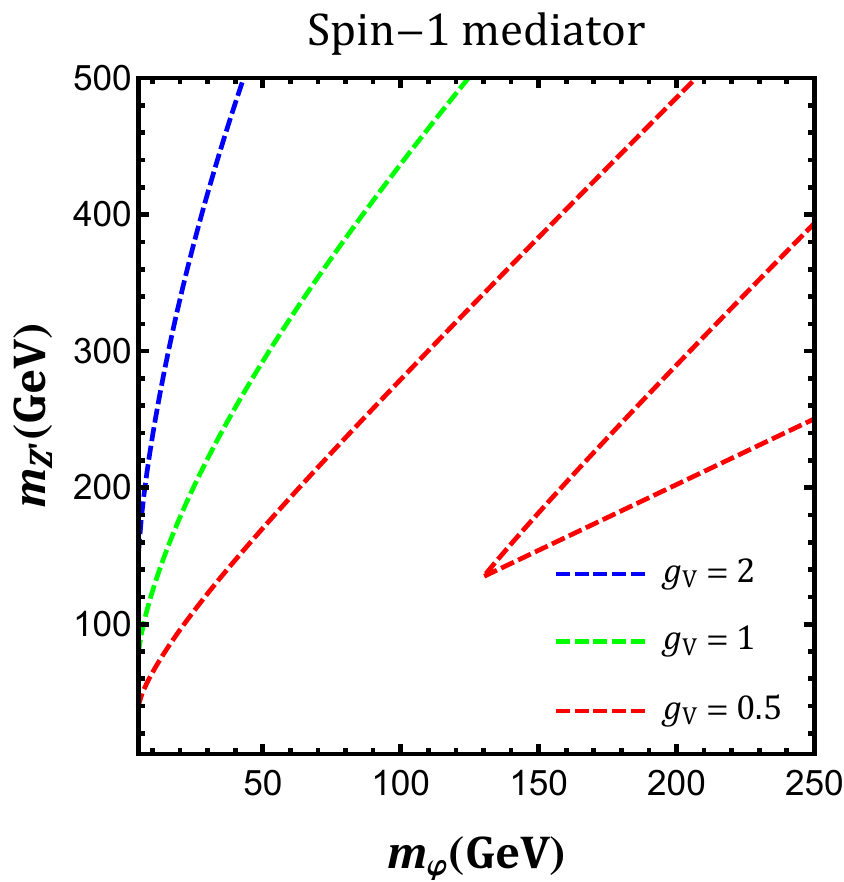}
\caption{Left panel: Contours of $\Lambda$ in the $m_{\varphi}-m_{\Phi}$ plane by setting $f_{P}=1.5$, $\lambda=0.1$. Right panel: Contours of $g_{V}$ in the $m_{\varphi}-m_{Z'}$ plane by setting $\zeta = g_{V}$.}
\label{fig:contour}
\end{figure}

We show in the left panel of Fig.~\ref{fig:contour} contour plot of $\Lambda$ in the $m_{\varphi}-m_{\Phi}$ plane, by setting $f_P = 1.5$, $\lambda = 0.1$. The red, green and blue dashed lines correspond to $\Lambda = 0.5,\, 1, \, 2$, respectively. There are  two gray solid lines in the plot, the lighter one for $m_{\Phi} = 2m_{\varphi}$ and the darker one for $m_{\Phi} = m_{\varphi}$, which divide the $m_{\varphi}-m_{\Phi}$ plane into three regions, I, II and III. For the region I and II, there is only one annihilation channel $\varphi^\dag\varphi \rightarrow \bar{b}b$ allowed, where the only difference is that the decay channel $\Phi \rightarrow \varphi^\dag\varphi$ is allowed in the region I, but forbidden in the region II. In the region III, the new channel $\varphi^\dag\varphi \rightarrow \Phi\Phi$ is allowed. The contours for $g_{V}$ are shown in the right panel of Fig.~\ref{fig:contour}, where the red, green and blue dashed lines correspond to $g_V=0.5,~1,~2$ respectively. Notice that $\mathcal{O}(g_{V}) \sim 1$ for $m_{Z'} \sim m_{\varphi} \sim 200\, {\rm GeV}$.

\begin{figure}[t]
\centering
\includegraphics[scale=0.75]{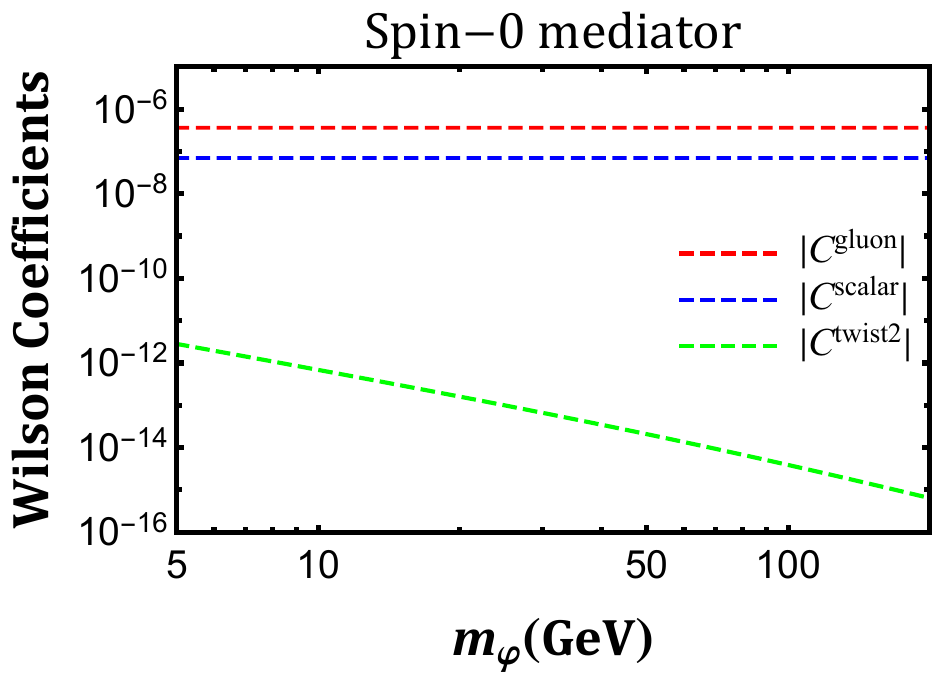}
\hspace{0.5cm}
\includegraphics[scale=0.75]{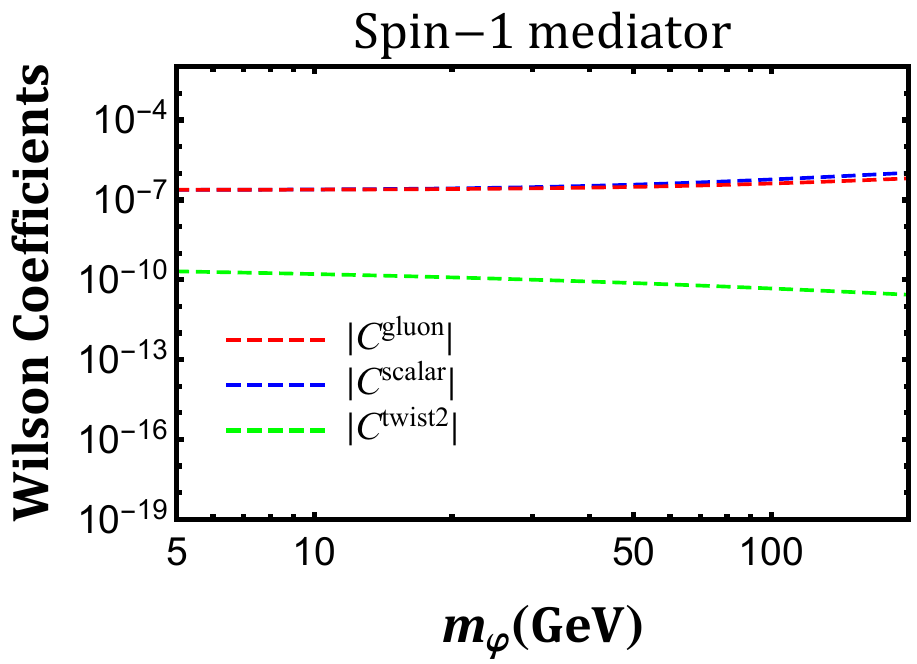}
\caption{Wilson coefficients as the function of the dark matter mass $m_{\varphi}$ by setting $f_{P}=1.5$, $\lambda=0.1$, $\Lambda=0.5$, $m_{\Phi}=100\,{\rm GeV}$ for spin-0 mediator (left-panel) and $\zeta=g_{V}=0.5$, $m_{Z'}=100\,{\rm GeV}$ for spin-1 mediator (right-panel).}
\label{fig:wilson}
\end{figure}

We show in the Fig.~\ref{fig:wilson} Wilson coefficients as the function of the dark matter mass $m_{\varphi}$. The plot for the spin-0 mediator case is given in the left-panel, by setting $f_{P}=1.5, \, \zeta=0.1, \, \Lambda=0.5$ and $m_{\Phi}=100\,{\rm GeV}$. As can be seen, the Wilson coefficient for the gluon-DM operator is the largest, which is of the order $\mathcal{O}(10^{-7})$. The Wilson coefficient for quark-DM operator is smaller than that of the gluon-DM operator by a factor  of $\mathcal{O}(5)$. The plot in the right panel shows the Wilson coefficients for spin-1 mediator case by setting $\zeta=g_V=0.5$. Notice that the Wilson coefficients of the gluon-DM and the scalar-type quark-DM operators are almost the same.

\begin{figure}[htbp]
\centering
\includegraphics[scale=0.75]{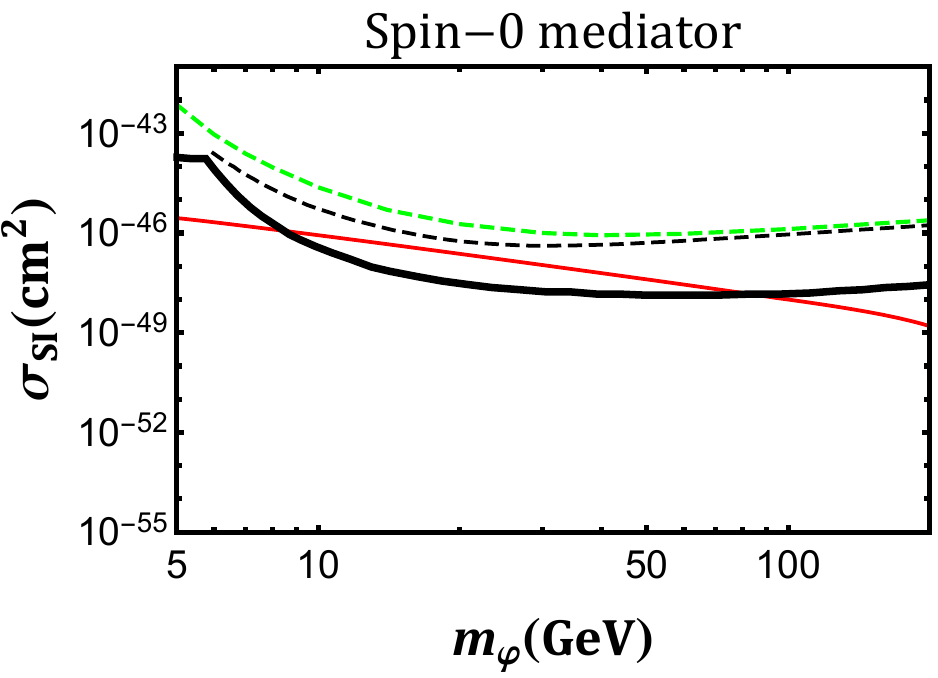}
\hspace{0.5cm}
\includegraphics[scale=0.75]{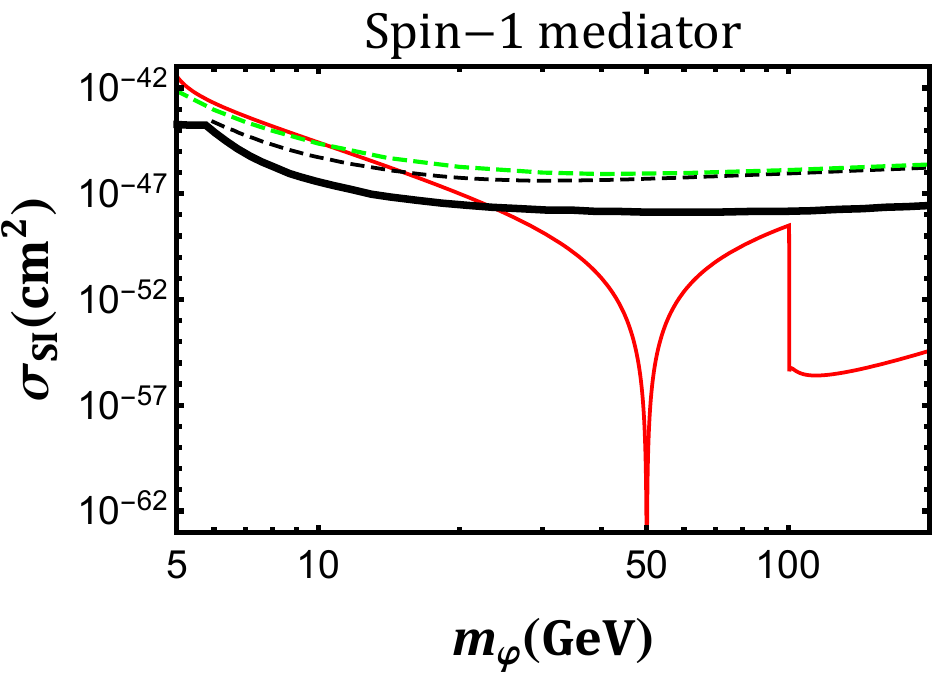}
\caption{Spin-independent DM-nucleon cross section as the function of the dark matter mass $m_{\varphi}$ by setting $f_{P}=1.5$, $\lambda=0.1$, $m_{\Phi}=100\,{\rm GeV}$ for spin-0 mediator (left-panel) and $\zeta=g_{V}$, $m_{Z'}=100\,{\rm GeV}$ for spin-1 mediator (right-panel). The green dashed line and black dashed line are constraints from XENON1T and PandaX-II experiments, respectively. The black solid line denotes the neutrino floor.}
\label{fig:cross section}
\end{figure}

Finally, we consider the spin-independent DM-nucleon elastic scattering cross section. As an illustration, we show in the Fig.~\ref{fig:cross section} the direct detection cross section as the function of the dark matter mass $m_{\varphi}$. The spin-0 mediator case is shown in the left-panel by setting $f_{P}=1.5, \lambda=0.1$ and $m_{\Phi}=100\,{\rm GeV}$. The green dashed and black dashed lines are exclusion limits put by XENON1T~\cite{Aprile:2017iyp} and PandaX-II~\cite{Tan:2016zwf} experiments,  separately. There is a wide range of mass $(9 - 80)\,{\rm GeV}$ that can be detected with the help of current direct detection techniques. The plot in the right-panel corresponds to the spin-1 mediator case by setting $\zeta=g_V$ and $M_{Z^\prime}=100~{\rm GeV}$, for which there is only a narrow mass range $(15 - 23)\,{\rm GeV}$ lying above the neutrino floor. The reason leading to a small cross section is that the contributions from quark-DM and gluon-DM operators cancels with each other as can be seen from the right-panel of the Fig.~\ref{fig:wilson}.

\section{Summary}

There are more than twenty DM direct detection experiments on the Earth searching for nuclear recoils induced by elastic WIMP-nucleon scattering and there are countless WIMP models on the market. Theoretically, which kind of WIMP deserves a deep investigation is a question worth pondering. It looks like to be a good strategy to study the direct detection signal of a WIMP model that has  non-trivial signals in indirect  detection experiments.  The bottom quark flavored DM  is one of  well motivated models since it can interpret the cosmic ray antiproton excess observed by the AMS-02 Collaboration as well as the GeV-scale gamma ray excess observed from the Galactic center. In this paper, we have studied next-to-leading order corrections to the scattering cross section of the scalar-type bottom-quark-flavored DM with nucleon.  We focused on two scenarios in which the direct detection cross sections are suppressed at the tree-level. Our results given in the Fig.~\ref{fig:cross section} show that next-to-leading order corrections are sizable and it is possible to detect some parameter space of such models using current direct detection techniques.
It should be mentioned that no matter the observed excesses in indirect detection experiments are induced by WIMP or not, an systematic study to the direct detection signal of a WIMP that might be detected in an indirect detection experiment is necessary as we do not know what kind of exotic signal will be brought to us in future observations.

\acknowledgments

This work was supported by the National Natural Science Foundation of China under grant No. 11775025 and the Fundamental Research Funds for the Central Universities under grant No. 2017NT17.


\appendix
\section{Annihilation cross sections}\label{annihilation}
Annihilation cross sections corresponding to spin-0 and spin-1 mediators are listed respectively as follows 
\begin{itemize}
\item{spin-0 mediator ($m_\varphi < m_\Phi$)}
\begin{eqnarray}
\sigma v = && \frac{f_P^2 \Lambda ^2 \sqrt{m_\varphi^2-m_b^2}}{4 \pi m_\varphi \big((m_\Phi^2-4 m_\varphi^2)^2+\Gamma^2_{\Phi}m_{\Phi}^2\big)} \nonumber \\
           && + \frac{v^2 f_P^2\Lambda^2}{32\pi m_\varphi\sqrt{m_\varphi^2-m_b^2}\big((m_\Phi^2-4m_\varphi^2)^2 + \Gamma_{\Phi}^2m_{\Phi}^2\big)^2} \Big[ -2m_\varphi^2 m_\Phi^2(\Gamma_{\Phi}^2 + m_{\Phi}^2 + 20m_b^2) \nonumber \\
           && +16m_{\varphi}^4(2m_{\Phi}^2 + 7m_b^2) - 96m_{\varphi}^6 +3m_{\Phi}^2m_{b}^2(\Gamma_{\Phi}^2+m_{\Phi}^2) \Big]
\end{eqnarray}
\item{spin-0 mediator ($m_\varphi > m_\Phi$)}
\begin{eqnarray}
\sigma v = && \frac{\sqrt{m_\varphi^2 - m_\Phi^2}}{64\pi m_\varphi^3(2m_\varphi^2 - m_\Phi^2)^2}(2\Lambda^2 - 2\lambda m_\varphi^2 + \lambda m_{\Phi}^2)^2 \nonumber \\
           && + \frac{v^2 (2\Lambda^2 - 2\lambda m_{\varphi}^2 + \lambda m_{\Phi}^2)}{1536\pi m_\varphi^3\sqrt{m_\varphi^2 - m_\Phi^2}(2m_\varphi^2-m_\Phi^2)^4}\Big[3\lambda(4m_{\varphi}^2 - 5m_{\Phi}^2)(2m_{\varphi}^2 - m_{\Phi}^2)^3 \nonumber \\
           && + 2\Lambda^2(148m_{\varphi}^4m_{\Phi}^2 - 80m_{\varphi}^2m_{\Phi}^4 - 80 m_{\varphi}^6 + 15m_{\Phi}^6) \Big]
\end{eqnarray}
\item{spin-1 mediator ($m_\varphi < m_{Z'}$)}
\begin{eqnarray}
\sigma v = \frac{v^2\zeta^2g_V^2(2m_\varphi^2+m_b^2)}{12\pi\big[(m_{Z'}^2-4m_\varphi^2)^2+m_{Z'}^2\Gamma_{Z'}^2\big]}\sqrt{1-\frac{m_b^2}{m_\varphi^2}}
\end{eqnarray}
\item{spin-1 mediator ($m_\varphi > m_{Z'}$)}
\begin{eqnarray}
\sigma v = && \frac{g_V^4\sqrt{1-m_{Z'}^2/m_{\varphi}^2}}{16\pi m_{\varphi}^2(2m_{\varphi}^2-m_{Z'}^2)^2}( -8m_{\varphi}^2m_{Z'}^2 + 8m_{\varphi}^4 + 3m_{Z'}^4 ) \nonumber \\
          && + \frac{v^2g_V^4}{384\pi m_{\varphi}^4\sqrt{1-m_{Z'}^2/m_{\varphi}^2}(2m_{\varphi}^2-m_{Z'}^2)^4}\Big( 1888m_{\varphi}^8m_{Z'}^2 - 2224m_{\varphi}^6m_{Z'}^4 \nonumber \\
          && + 1332m_{\varphi}^4m_{Z'}^6 - 392m_{\varphi}^2m_{Z'}^8 - 640m_{\varphi}^{10} + 45m_{Z'}^{10} \Big)
\end{eqnarray}
\end{itemize}

\section{Loop functions}\label{loop}
Loop functions used in this work are collected as follows \cite{Passarino:1978jh, Abe:2015rja}:
\begin{eqnarray}
\int{d^4k \over (2\pi)^4}{1 \over [(p+k)^2-m_\varphi^2][k^2-m_a^2]} &=& {i \over 16\pi^2}B_0(p^2,m_a^2,m_\varphi^2) \\
\int{d^4k \over (2\pi)^4}{1 \over [(p+k)^2-m_\varphi^2][k^2-m_a^2]^2} &=& {i \over 16\pi^2}C_0(p^2,m_a^2,m_\varphi^2) \\
\int {d^4k \over (2\pi)^4}{1 \over [(p+k)^2-m_\varphi^2][k^2-m_a^2]^3} &=& {i \over 16\pi^2}D_0(p^2,m_a^2,m_\varphi^2) \\
\int{d^4k \over (2\pi)^4}{k^\mu \over [(p+k)^2-m_\varphi^2][k^2-m_a^2]^2} &=& {i \over 16\pi^2}p^\mu C_2(p^2,m_a^2,m_\varphi^2) \\
\int{d^4k \over (2\pi)^4}{1 \over [(p+k)^2-m_\varphi^2][k^2-m_a^2][k^2-m_Q^2]^n} &=& {i \over 16\pi^2}X_n(p^2,m_\varphi^2,m_a^2,m_Q^2) \\
\int{d^4k \over (2\pi)^4}{k^{\mu} \over [(p+k)^2-m_\varphi^2][k^2-m_a^2][k^2-m_Q^2]^n} &=& {i \over 16\pi^2}p^{\mu}Y_n(p^2,m_\varphi^2,m_a^2,m_Q^2) \\
\int{d^4k \over (2\pi)^4}{k^\mu k^\nu \over [(p+k)^2-m_\varphi^2]k^4[k^2-m_a^2]^2} &=& {i \over 16\pi^2}[p^\mu p^\nu Z_{11}(p^2,m_\varphi^2,m_a^2) \nonumber\\ && + g^{\mu\nu}Z_{00}(p^2,m_\varphi^2,m_a^2)]
\end{eqnarray}
The expressions for $\partial F_1(m_a^2)/\partial(m_a^2)$ and $\partial F_2(m_a^2)/\partial(m_a^2)$ using the loop functions are
\begin{eqnarray}
{\partial \over \partial m_a^2} F_1(m_a^2) &=& \int_0^1 dx \Big\{3 {\partial \over \partial m_a^2} X_1(m_\varphi^2, m_\varphi^2, m_a^2, {m_Q^2 \over x(1-x)}) \nonumber \\
& & - m_Q^2 \frac{(2+5x-5x^2)}{x^2(1-x)^2}{\partial \over \partial m_a^2} X_2(m_\varphi^2, m_\varphi^2, m_a^2, {m_Q^2 \over x(1-x)}) \nonumber \\ 
& & - 2m_Q^4 \frac{(1-2x+2x^2)}{x^3(1-x)^3}{\partial \over \partial m_a^2} X_3(m_\varphi^2, m_\varphi^2, m_a^2, {m_Q^2 \over x(1-x)})\Big\} \\
{\partial \over \partial m_a^2} F_2(m_a^2) &=& \int_0^1 dx \Big\{3 {\partial \over \partial m_a^2} B_0(0, {m_Q^2 \over x(1-x)}, m_a^2) \nonumber \\
& & - m_Q^2 \frac{(2+5x-5x^2)}{x^2(1-x)^2} {\partial \over \partial m_a^2} C_0(0, {m_Q^2 \over x(1-x)}, m_a^2) \nonumber \\ 
& & - 2m_Q^4 \frac{(1-2x+2x^2)}{x^3(1-x)^3} {\partial \over \partial m_a^2} D_0(0, {m_Q^2 \over x(1-x)}, m_a^2)\Big\}
\end{eqnarray}
All of these functions can be evaluated by using Package-X \cite{Patel:2015tea,Patel:2016fam}. For example, we give the analytical expression for $C_0(p^2,m_a^2,m_\varphi^2)$,
\begin{eqnarray}
C_0(p^2,m_a^2,m_\varphi^2) &=& \frac{1}{2p^2}\log\Big({m_\varphi^2 \over m_a^2}\Big) - \Bigg[\frac{(m_\varphi^2-m_a^2+p^2)}{p^2\sqrt{m_a^4-2m_a^2m_\varphi^2-2m_a^2p^2+m_\varphi^4-2m_\varphi^2p^2+p^4}} \nonumber \\
& & \times \log\Big(\frac{m_a^2+m_\varphi^2-p^2+\sqrt{m_a^4-2m_a^2m_\varphi^2-2m_a^2p^2+m_\varphi^4-2m_\varphi^2p^2+p^4}}{2m_a m_\varphi} \Big) \Bigg]  \nonumber \\
\end{eqnarray}
and the remaining functions can be evaluated in the same way.

\section{Nucleon form factors}\label{nucleon matrix elements}
We here give the definition of the nuclear form factors required to calculate the effective interactions between DM and nucleons. For the spin-independent interactions, we need the following nuclear form factors \cite{Shifman:1978zn, Jungman:1995df}:
\begin{eqnarray}
\langle N| m_q \bar{q}q |N \rangle &=& m_N f_{T_q}^N, \quad (q=u,d,s) \\
\langle N|-\frac{9\alpha_s}{8\pi}G^a_{\mu\nu}G^{a\mu\nu} |N \rangle &=& m_N f_{T_G}^N, \\
\langle N|{\cal O}_{\mu\nu}^q |N \rangle &=& \frac{1}{m_N}(p^{N}_\mu p^{N}_\nu - {1 \over 4} m_N^2 g_{\mu\nu})(q^N(2) + \bar{q}^N(2))
\end{eqnarray}
where $m_N$ is the nucleon mass, $f_{T_q}^N$ and $f_{T_G}^N$ are nuclear form factors, $q^N$ and $\bar q^N$ are the second moments of the quark parton distribution functions and $p^{N}_\mu$ is the nucleon four-momentum. The present numerical values of the form factors for light quarks are taken from micrOMEGAs \cite{Belanger:2008sj, Belanger:2018ccd}
\begin{eqnarray}
\begin{aligned}
f_{T_u}^{p} &= 0.01513, \qquad  & f_{T_d}^{p} = 0.0191, \qquad  f_{T_s}^p = 0.0447, \\
f_{T_u}^n &= 0.0110,  \qquad  & f_{T_d}^n = 0.0273, \qquad  f_{T_s}^n = 0.0447,
\end{aligned}
\end{eqnarray}
which can be related to the gluon form factor via \cite{Jungman:1995df}
\begin{eqnarray}
f_{T_G}^{p(n)} = 1 - \sum_{q=u,d,s} f_{T_q}^{p(n)}.
\end{eqnarray}
The second moments are calculated at the scale $\mu=m_{Z}$ by using CTEQ parton distribution functions \cite{Pumplin:2002vw,Abe:2018emu}
\begin{eqnarray}
\begin{aligned}
g^{p}(2) &= 0.464, \\
u^p(2) &= 0.22,  \qquad  & \bar{u}^p(2) = 0.034, \\
d^p(2) &= 0.11,  \qquad  & \bar{d}^p(2) = 0.036, \\
s^p(2) &= 0.026,  \qquad & \bar{s}^p(2) = 0.026, \\
c^p(2) &= 0.019,  \qquad & \bar{c}^p(2) = 0.019, \\
b^p(2) &= 0.012,  \qquad & \bar{b}^p(2) = 0.012,
\end{aligned}
\end{eqnarray}
where the corresponding values for neutron can be obtained by interchanging up and down quarks.


\bibliographystyle{JHEP}
\bibliography{reference.bib}

\providecommand{\href}[2]{#2}\begingroup\raggedright\begin{thebibliography}{10}

\bibitem{Aghanim:2018eyx}
{\scshape Planck} collaboration, \emph{{Planck 2018 results. VI. Cosmological
  parameters}},
  \href{https://doi.org/10.1051/0004-6361/201833910}{\emph{Astron. Astrophys.}
  {\bfseries 641} (2020) A6}
  [\href{https://arxiv.org/abs/1807.06209}{{\ttfamily 1807.06209}}].

\bibitem{Goldberg:1983nd}
H.~Goldberg, \emph{{Constraint on the Photino Mass from Cosmology}},
  \href{https://doi.org/10.1103/PhysRevLett.103.099905,
  10.1103/PhysRevLett.50.1419}{\emph{Phys. Rev. Lett.} {\bfseries 50} (1983)
  1419}.

\bibitem{Ellis:1983ew}
J.R.~Ellis, J.S.~Hagelin, D.V.~Nanopoulos, K.A.~Olive and M.~Srednicki,
  \emph{{Supersymmetric Relics from the Big Bang}},
  \href{https://doi.org/10.1016/0550-3213(84)90461-9}{\emph{Nucl. Phys.}
  {\bfseries B238} (1984) 453}.

\bibitem{Jungman:1995df}
G.~Jungman, M.~Kamionkowski and K.~Griest, \emph{{Supersymmetric dark matter}},
  \href{https://doi.org/10.1016/0370-1573(95)00058-5}{\emph{Phys. Rept.}
  {\bfseries 267} (1996) 195}
  [\href{https://arxiv.org/abs/hep-ph/9506380}{{\ttfamily hep-ph/9506380}}].

\bibitem{Servant:2002aq}
G.~Servant and T.M.P.~Tait, \emph{{Is the lightest Kaluza-Klein particle a
  viable dark matter candidate?}},
  \href{https://doi.org/10.1016/S0550-3213(02)01012-X}{\emph{Nucl. Phys.}
  {\bfseries B650} (2003) 391}
  [\href{https://arxiv.org/abs/hep-ph/0206071}{{\ttfamily hep-ph/0206071}}].

\bibitem{Cheng:2002ej}
H.-C.~Cheng, J.L.~Feng and K.T.~Matchev, \emph{{Kaluza-Klein dark matter}},
  \href{https://doi.org/10.1103/PhysRevLett.89.211301}{\emph{Phys. Rev. Lett.}
  {\bfseries 89} (2002) 211301}
  [\href{https://arxiv.org/abs/hep-ph/0207125}{{\ttfamily hep-ph/0207125}}].

\bibitem{Bertone:2004pz}
G.~Bertone, D.~Hooper and J.~Silk, \emph{{Particle dark matter: Evidence,
  candidates and constraints}},
  \href{https://doi.org/10.1016/j.physrep.2004.08.031}{\emph{Phys. Rept.}
  {\bfseries 405} (2005) 279}
  [\href{https://arxiv.org/abs/hep-ph/0404175}{{\ttfamily hep-ph/0404175}}].

\bibitem{Lin:2019uvt}
T.~Lin, \emph{{Dark matter models and direct detection}},
  \href{https://doi.org/10.22323/1.333.0009}{\emph{PoS} {\bfseries 333} (2019)
  009} [\href{https://arxiv.org/abs/1904.07915}{{\ttfamily 1904.07915}}].

\bibitem{Slatyer:2017sev}
T.R.~Slatyer, \emph{{Indirect Detection of Dark Matter}},  in
  \emph{{Proceedings, Theoretical Advanced Study Institute in Elementary
  Particle Physics : Anticipating the Next Discoveries in Particle Physics
  (TASI 2016): Boulder, CO, USA, June 6-July 1, 2016}}, pp.~297--353, 2018,
  \href{https://doi.org/10.1142/9789813233348_0005}{DOI}
  [\href{https://arxiv.org/abs/1710.05137}{{\ttfamily 1710.05137}}].

\bibitem{Hooper:2018kfv}
D.~Hooper, \emph{{TASI Lectures on Indirect Searches For Dark Matter}},
  {\emph{PoS} {\bfseries TASI2018} (2019) 010}
  [\href{https://arxiv.org/abs/1812.02029}{{\ttfamily 1812.02029}}].

\bibitem{Aguilar:2016kjl}
{\scshape AMS} collaboration, \emph{{Antiproton Flux, Antiproton-to-Proton Flux
  Ratio, and Properties of Elementary Particle Fluxes in Primary Cosmic Rays
  Measured with the Alpha Magnetic Spectrometer on the International Space
  Station}}, \href{https://doi.org/10.1103/PhysRevLett.117.091103}{\emph{Phys.
  Rev. Lett.} {\bfseries 117} (2016) 091103}.

\bibitem{Cholis:2019ejx}
I.~Cholis, T.~Linden and D.~Hooper, \emph{{A Robust Excess in the Cosmic-Ray
  Antiproton Spectrum: Implications for Annihilating Dark Matter}},
  \href{https://doi.org/10.1103/PhysRevD.99.103026}{\emph{Phys. Rev.}
  {\bfseries D99} (2019) 103026}
  [\href{https://arxiv.org/abs/1903.02549}{{\ttfamily 1903.02549}}].

\bibitem{Goodenough:2009gk}
L.~Goodenough and D.~Hooper, \emph{{Possible Evidence For Dark Matter
  Annihilation In The Inner Milky Way From The Fermi Gamma Ray Space
  Telescope}},  \href{https://arxiv.org/abs/0910.2998}{{\ttfamily 0910.2998}}.

\bibitem{Hooper:2010mq}
D.~Hooper and L.~Goodenough, \emph{{Dark Matter Annihilation in The Galactic
  Center As Seen by the Fermi Gamma Ray Space Telescope}},
  \href{https://doi.org/10.1016/j.physletb.2011.02.029}{\emph{Phys. Lett.}
  {\bfseries B697} (2011) 412}
  [\href{https://arxiv.org/abs/1010.2752}{{\ttfamily 1010.2752}}].

\bibitem{Hooper:2011ti}
D.~Hooper and T.~Linden, \emph{{On The Origin Of The Gamma Rays From The
  Galactic Center}},
  \href{https://doi.org/10.1103/PhysRevD.84.123005}{\emph{Phys. Rev.}
  {\bfseries D84} (2011) 123005}
  [\href{https://arxiv.org/abs/1110.0006}{{\ttfamily 1110.0006}}].

\bibitem{Abazajian:2012pn}
K.N.~Abazajian and M.~Kaplinghat, \emph{{Detection of a Gamma-Ray Source in the
  Galactic Center Consistent with Extended Emission from Dark Matter
  Annihilation and Concentrated Astrophysical Emission}},
  \href{https://doi.org/10.1103/PhysRevD.86.083511,
  10.1103/PhysRevD.87.129902}{\emph{Phys. Rev.} {\bfseries D86} (2012) 083511}
  [\href{https://arxiv.org/abs/1207.6047}{{\ttfamily 1207.6047}}].

\bibitem{Daylan:2014rsa}
T.~Daylan, D.P.~Finkbeiner, D.~Hooper, T.~Linden, S.K.N.~Portillo, N.L.~Rodd
  et~al., \emph{{The characterization of the gamma-ray signal from the central
  Milky Way: A case for annihilating dark matter}},
  \href{https://doi.org/10.1016/j.dark.2015.12.005}{\emph{Phys. Dark Univ.}
  {\bfseries 12} (2016) 1} [\href{https://arxiv.org/abs/1402.6703}{{\ttfamily
  1402.6703}}].

\bibitem{Hooper:2019xss}
D.~Hooper, R.K.~Leane, Y.-D.~Tsai, S.~Wegsman and S.J.~Witte, \emph{{A
  systematic study of hidden sector dark matter:application to the gamma-ray
  and antiproton excesses}},
  \href{https://doi.org/10.1007/JHEP07(2020)163}{\emph{JHEP} {\bfseries 07}
  (2020) 163} [\href{https://arxiv.org/abs/1912.08821}{{\ttfamily
  1912.08821}}].

\bibitem{Berlin:2014tja}
A.~Berlin, D.~Hooper and S.D.~McDermott, \emph{{Simplified Dark Matter Models
  for the Galactic Center Gamma-Ray Excess}},
  \href{https://doi.org/10.1103/PhysRevD.89.115022}{\emph{Phys. Rev.}
  {\bfseries D89} (2014) 115022}
  [\href{https://arxiv.org/abs/1404.0022}{{\ttfamily 1404.0022}}].

\bibitem{Haisch:2013uaa}
U.~Haisch and F.~Kahlhoefer, \emph{{On the importance of loop-induced
  spin-independent interactions for dark matter direct detection}},
  \href{https://doi.org/10.1088/1475-7516/2013/04/050}{\emph{JCAP} {\bfseries
  1304} (2013) 050} [\href{https://arxiv.org/abs/1302.4454}{{\ttfamily
  1302.4454}}].

\bibitem{Crivellin:2014gpa}
A.~Crivellin and U.~Haisch, \emph{{Dark matter direct detection constraints
  from gauge bosons loops}},
  \href{https://doi.org/10.1103/PhysRevD.90.115011}{\emph{Phys. Rev.}
  {\bfseries D90} (2014) 115011}
  [\href{https://arxiv.org/abs/1408.5046}{{\ttfamily 1408.5046}}].

\bibitem{DEramo:2016gos}
F.~D'Eramo, B.J.~Kavanagh and P.~Panci, \emph{{You can hide but you have to
  run: direct detection with vector mediators}},
  \href{https://doi.org/10.1007/JHEP08(2016)111}{\emph{JHEP} {\bfseries 08}
  (2016) 111} [\href{https://arxiv.org/abs/1605.04917}{{\ttfamily
  1605.04917}}].

\bibitem{Crivellin:2014qxa}
A.~Crivellin, F.~D'Eramo and M.~Procura, \emph{{New Constraints on Dark Matter
  Effective Theories from Standard Model Loops}},
  \href{https://doi.org/10.1103/PhysRevLett.112.191304}{\emph{Phys. Rev. Lett.}
  {\bfseries 112} (2014) 191304}
  [\href{https://arxiv.org/abs/1402.1173}{{\ttfamily 1402.1173}}].

\bibitem{Bishara:2018vix}
F.~Bishara, J.~Brod, B.~Grinstein and J.~Zupan, \emph{{Renormalization Group
  Effects in Dark Matter Interactions}},
  \href{https://doi.org/10.1007/JHEP03(2020)089}{\emph{JHEP} {\bfseries 03}
  (2020) 089} [\href{https://arxiv.org/abs/1809.03506}{{\ttfamily
  1809.03506}}].

\bibitem{Li:2018qip}
T.~Li, \emph{{Revisiting the direct detection of dark matter in simplified
  models}}, \href{https://doi.org/10.1016/j.physletb.2018.05.073}{\emph{Phys.
  Lett.} {\bfseries B782} (2018) 497}
  [\href{https://arxiv.org/abs/1804.02120}{{\ttfamily 1804.02120}}].

\bibitem{Sanderson:2018lmj}
N.F.~Bell, G.~Busoni and I.W.~Sanderson, \emph{{Loop Effects in Direct
  Detection}}, \href{https://doi.org/10.1088/1475-7516/2018/08/017,
  10.1088/1475-7516/2019/01/E01}{\emph{JCAP} {\bfseries 1808} (2018) 017}
  [\href{https://arxiv.org/abs/1803.01574}{{\ttfamily 1803.01574}}].

\bibitem{Hisano:2010fy}
J.~Hisano, K.~Ishiwata and N.~Nagata, \emph{{A complete calculation for direct
  detection of Wino dark matter}},
  \href{https://doi.org/10.1016/j.physletb.2010.05.047}{\emph{Phys. Lett.}
  {\bfseries B690} (2010) 311}
  [\href{https://arxiv.org/abs/1004.4090}{{\ttfamily 1004.4090}}].

\bibitem{Hisano:2011cs}
J.~Hisano, K.~Ishiwata, N.~Nagata and T.~Takesako, \emph{{Direct Detection of
  Electroweak-Interacting Dark Matter}},
  \href{https://doi.org/10.1007/JHEP07(2011)005}{\emph{JHEP} {\bfseries 07}
  (2011) 005} [\href{https://arxiv.org/abs/1104.0228}{{\ttfamily 1104.0228}}].

\bibitem{Ertas:2019dew}
F.~Ertas and F.~Kahlhoefer, \emph{{Loop-induced direct detection signatures
  from CP-violating scalar mediators}},
  \href{https://doi.org/10.1007/JHEP06(2019)052}{\emph{JHEP} {\bfseries 06}
  (2019) 052} [\href{https://arxiv.org/abs/1902.11070}{{\ttfamily
  1902.11070}}].

\bibitem{Ishiwata:2018sdi}
K.~Ishiwata and T.~Toma, \emph{{Probing pseudo Nambu-Goldstone boson dark
  matter at loop level}},
  \href{https://doi.org/10.1007/JHEP12(2018)089}{\emph{JHEP} {\bfseries 12}
  (2018) 089} [\href{https://arxiv.org/abs/1810.08139}{{\ttfamily
  1810.08139}}].

\bibitem{Abe:2018emu}
T.~Abe, M.~Fujiwara and J.~Hisano, \emph{{Loop corrections to dark matter
  direct detection in a pseudoscalar mediator dark matter model}},
  \href{https://doi.org/10.1007/JHEP02(2019)028}{\emph{JHEP} {\bfseries 02}
  (2019) 028} [\href{https://arxiv.org/abs/1810.01039}{{\ttfamily
  1810.01039}}].

\bibitem{Chao:2018xwz}
W.~Chao, G.-J.~Ding, X.-G.~He and M.~Ramsey-Musolf, \emph{{Scalar Electroweak
  Multiplet Dark Matter}},
  \href{https://doi.org/10.1007/JHEP08(2019)058}{\emph{JHEP} {\bfseries 08}
  (2019) 058} [\href{https://arxiv.org/abs/1812.07829}{{\ttfamily
  1812.07829}}].

\bibitem{Ghorbani:2018pjh}
K.~Ghorbani and P.H.~Ghorbani, \emph{{Leading Loop Effects in
  Pseudoscalar-Higgs Portal Dark Matter}},
  \href{https://doi.org/10.1007/JHEP05(2019)096}{\emph{JHEP} {\bfseries 05}
  (2019) 096} [\href{https://arxiv.org/abs/1812.04092}{{\ttfamily
  1812.04092}}].

\bibitem{Li:2019fnn}
T.~Li and P.~Wu, \emph{{Simplified dark matter models with loop effects in
  direct detection and the constraints from indirect detection and collider
  search}}, \href{https://doi.org/10.1088/1674-1137/43/11/113102}{\emph{Chin.
  Phys.} {\bfseries C43} (2019) 113102}
  [\href{https://arxiv.org/abs/1904.03407}{{\ttfamily 1904.03407}}].

\bibitem{Chao:2019lhb}
W.~Chao, \emph{{Direct detections of Majorana dark matter in vector portal}},
  \href{https://doi.org/10.1007/JHEP11(2019)013}{\emph{JHEP} {\bfseries 11}
  (2019) 013} [\href{https://arxiv.org/abs/1904.09785}{{\ttfamily
  1904.09785}}].

\bibitem{Monroe:2007xp}
J.~Monroe and P.~Fisher, \emph{{Neutrino Backgrounds to Dark Matter Searches}},
  \href{https://doi.org/10.1103/PhysRevD.76.033007}{\emph{Phys. Rev.}
  {\bfseries D76} (2007) 033007}
  [\href{https://arxiv.org/abs/0706.3019}{{\ttfamily 0706.3019}}].

\bibitem{Strigari:2009bq}
L.E.~Strigari, \emph{{Neutrino Coherent Scattering Rates at Direct Dark Matter
  Detectors}}, \href{https://doi.org/10.1088/1367-2630/11/10/105011}{\emph{New
  J. Phys.} {\bfseries 11} (2009) 105011}
  [\href{https://arxiv.org/abs/0903.3630}{{\ttfamily 0903.3630}}].

\bibitem{Billard:2013qya}
J.~Billard, L.~Strigari and E.~Figueroa-Feliciano, \emph{{Implication of
  neutrino backgrounds on the reach of next generation dark matter direct
  detection experiments}},
  \href{https://doi.org/10.1103/PhysRevD.89.023524}{\emph{Phys. Rev.}
  {\bfseries D89} (2014) 023524}
  [\href{https://arxiv.org/abs/1307.5458}{{\ttfamily 1307.5458}}].

\bibitem{Gelmini:2018ogy}
G.B.~Gelmini, V.~Takhistov and S.J.~Witte, \emph{{Casting a Wide Signal Net
  with Future Direct Dark Matter Detection Experiments}},
  \href{https://doi.org/10.1088/1475-7516/2018/07/009,
  10.1088/1475-7516/2019/02/E02}{\emph{JCAP} {\bfseries 1807} (2018) 009}
  [\href{https://arxiv.org/abs/1804.01638}{{\ttfamily 1804.01638}}].

\bibitem{Boehm:2018sux}
C.~Boehm, D.G.~Cerdeno, P.A.N.~Machado, A.~Olivares-Del~Campo, E.~Perdomo and
  E.~Reid, \emph{{How high is the neutrino floor?}},
  \href{https://doi.org/10.1088/1475-7516/2019/01/043}{\emph{JCAP} {\bfseries
  1901} (2019) 043} [\href{https://arxiv.org/abs/1809.06385}{{\ttfamily
  1809.06385}}].

\bibitem{Chao:2019pyh}
W.~Chao, J.-G.~Jiang, X.~Wang and X.-Y.~Zhang, \emph{{Direct Detections of Dark
  Matter in the Presence of Non-standard Neutrino Interactions}},
  \href{https://doi.org/10.1088/1475-7516/2019/08/010}{\emph{JCAP} {\bfseries
  1908} (2019) 010} [\href{https://arxiv.org/abs/1904.11214}{{\ttfamily
  1904.11214}}].

\bibitem{Hisano:2015bma}
J.~Hisano, R.~Nagai and N.~Nagata, \emph{{Effective Theories for Dark Matter
  Nucleon Scattering}},
  \href{https://doi.org/10.1007/JHEP05(2015)037}{\emph{JHEP} {\bfseries 05}
  (2015) 037} [\href{https://arxiv.org/abs/1502.02244}{{\ttfamily
  1502.02244}}].

\bibitem{Hisano:2017jmz}
J.~Hisano, \emph{{Effective theory approach to direct detection of dark
  matter}},  in \emph{{Proceedings, Les Houches summer school: EFT in Particle
  Physics and Cosmology: Les Houches (Chamonix Valley), France}}, vol.~108,
  2020, \href{https://doi.org/10.1093/oso/9780198855743.003.0011}{DOI}
  [\href{https://arxiv.org/abs/1712.02947}{{\ttfamily 1712.02947}}].

\bibitem{Hisano:2010ct}
J.~Hisano, K.~Ishiwata and N.~Nagata, \emph{{Gluon contribution to the dark
  matter direct detection}},
  \href{https://doi.org/10.1103/PhysRevD.82.115007}{\emph{Phys. Rev.}
  {\bfseries D82} (2010) 115007}
  [\href{https://arxiv.org/abs/1007.2601}{{\ttfamily 1007.2601}}].

\bibitem{Hisano:2015rsa}
J.~Hisano, K.~Ishiwata and N.~Nagata, \emph{{QCD Effects on Direct Detection of
  Wino Dark Matter}},
  \href{https://doi.org/10.1007/JHEP06(2015)097}{\emph{JHEP} {\bfseries 06}
  (2015) 097} [\href{https://arxiv.org/abs/1504.00915}{{\ttfamily
  1504.00915}}].

\bibitem{1991NuPhB.360..145G}
P.~{Gondolo} and G.~{Gelmini}, \emph{{Cosmic abundances of stable particles:
  improved analysis.}},
  \href{https://doi.org/10.1016/0550-3213(91)90438-4}{\emph{Nuclear Physics B}
  {\bfseries 360} (1991) 145}.

\bibitem{Ellis:2016jkw}
J.~Ellis, \emph{{TikZ-Feynman: Feynman diagrams with TikZ}},
  \href{https://doi.org/10.1016/j.cpc.2016.08.019}{\emph{Comput. Phys. Commun.}
  {\bfseries 210} (2017) 103}
  [\href{https://arxiv.org/abs/1601.05437}{{\ttfamily 1601.05437}}].

\bibitem{Lewin:1995rx}
J.~Lewin and P.~Smith, \emph{{Review of mathematics, numerical factors, and
  corrections for dark matter experiments based on elastic nuclear recoil}},
  \href{https://doi.org/10.1016/S0927-6505(96)00047-3}{\emph{Astropart. Phys.}
  {\bfseries 6} (1996) 87}.

\bibitem{Helm:1956zz}
R.H.~Helm, \emph{{Inelastic and Elastic Scattering of 187-Mev Electrons from
  Selected Even-Even Nuclei}},
  \href{https://doi.org/10.1103/PhysRev.104.1466}{\emph{Phys. Rev.} {\bfseries
  104} (1956) 1466}.

\bibitem{Read:2014qva}
J.~Read, \emph{{The Local Dark Matter Density}},
  \href{https://doi.org/10.1088/0954-3899/41/6/063101}{\emph{J. Phys. G}
  {\bfseries 41} (2014) 063101}
  [\href{https://arxiv.org/abs/1404.1938}{{\ttfamily 1404.1938}}].

\bibitem{Barger:2010gv}
V.~Barger, W.-Y.~Keung and D.~Marfatia, \emph{{Electromagnetic properties of
  dark matter: Dipole moments and charge form factor}},
  \href{https://doi.org/10.1016/j.physletb.2010.12.008}{\emph{Phys. Lett. B}
  {\bfseries 696} (2011) 74} [\href{https://arxiv.org/abs/1007.4345}{{\ttfamily
  1007.4345}}].

\bibitem{Shifman:1978zn}
M.A.~Shifman, A.~Vainshtein and V.I.~Zakharov, \emph{{Remarks on Higgs Boson
  Interactions with Nucleons}},
  \href{https://doi.org/10.1016/0370-2693(78)90481-1}{\emph{Phys. Lett. B}
  {\bfseries 78} (1978) 443}.

\bibitem{Ellis:2008hf}
J.R.~Ellis, K.A.~Olive and C.~Savage, \emph{{Hadronic Uncertainties in the
  Elastic Scattering of Supersymmetric Dark Matter}},
  \href{https://doi.org/10.1103/PhysRevD.77.065026}{\emph{Phys. Rev.}
  {\bfseries D77} (2008) 065026}
  [\href{https://arxiv.org/abs/0801.3656}{{\ttfamily 0801.3656}}].

\bibitem{Novikov:1983gd}
V.~Novikov, M.A.~Shifman, A.~Vainshtein and V.I.~Zakharov, \emph{{Calculations
  in External Fields in Quantum Chromodynamics. Technical Review}},
  {\emph{Fortsch. Phys.} {\bfseries 32} (1984) 585}.

\bibitem{Shtabovenko:2016sxi}
V.~Shtabovenko, R.~Mertig and F.~Orellana, \emph{{New Developments in FeynCalc
  9.0}}, \href{https://doi.org/10.1016/j.cpc.2016.06.008}{\emph{Comput. Phys.
  Commun.} {\bfseries 207} (2016) 432}
  [\href{https://arxiv.org/abs/1601.01167}{{\ttfamily 1601.01167}}].

\bibitem{Passarino:1978jh}
G.~Passarino and M.~Veltman, \emph{{One Loop Corrections for e+ e- Annihilation
  Into mu+ mu- in the Weinberg Model}},
  \href{https://doi.org/10.1016/0550-3213(79)90234-7}{\emph{Nucl. Phys. B}
  {\bfseries 160} (1979) 151}.

\bibitem{Abe:2015rja}
T.~Abe and R.~Sato, \emph{{Quantum corrections to the spin-independent cross
  section of the inert doublet dark matter}},
  \href{https://doi.org/10.1007/JHEP03(2015)109}{\emph{JHEP} {\bfseries 03}
  (2015) 109} [\href{https://arxiv.org/abs/1501.04161}{{\ttfamily
  1501.04161}}].

\bibitem{Chetyrkin:1997un}
K.G.~Chetyrkin, B.A.~Kniehl and M.~Steinhauser, \emph{{Decoupling relations to
  O (alpha-s**3) and their connection to low-energy theorems}},
  \href{https://doi.org/10.1016/S0550-3213(98)81004-3,
  10.1016/S0550-3213(97)00649-4}{\emph{Nucl. Phys.} {\bfseries B510} (1998) 61}
  [\href{https://arxiv.org/abs/hep-ph/9708255}{{\ttfamily hep-ph/9708255}}].

\bibitem{Gross:1974cs}
D.J.~Gross and F.~Wilczek, \emph{{ASYMPTOTICALLY FREE GAUGE THEORIES. 2.}},
  \href{https://doi.org/10.1103/PhysRevD.9.980}{\emph{Phys. Rev.} {\bfseries
  D9} (1974) 980}.

\bibitem{LopezHonorez:2006gr}
L.~Lopez~Honorez, E.~Nezri, J.F.~Oliver and M.H.G.~Tytgat, \emph{{The Inert
  Doublet Model: An Archetype for Dark Matter}},
  \href{https://doi.org/10.1088/1475-7516/2007/02/028}{\emph{JCAP} {\bfseries
  0702} (2007) 028} [\href{https://arxiv.org/abs/hep-ph/0612275}{{\ttfamily
  hep-ph/0612275}}].

\bibitem{Barger:2007im}
V.~Barger, P.~Langacker, M.~McCaskey, M.J.~Ramsey-Musolf and G.~Shaughnessy,
  \emph{{LHC Phenomenology of an Extended Standard Model with a Real Scalar
  Singlet}}, \href{https://doi.org/10.1103/PhysRevD.77.035005}{\emph{Phys.
  Rev.} {\bfseries D77} (2008) 035005}
  [\href{https://arxiv.org/abs/0706.4311}{{\ttfamily 0706.4311}}].

\bibitem{Mohapatra:1980qe}
R.N.~Mohapatra and R.E.~Marshak, \emph{{Local B-L Symmetry of Electroweak
  Interactions, Majorana Neutrinos and Neutron Oscillations}},
  \href{https://doi.org/10.1103/PhysRevLett.44.1644.2,
  10.1103/PhysRevLett.44.1316}{\emph{Phys. Rev. Lett.} {\bfseries 44} (1980)
  1316}.

\bibitem{Chao:2016avy}
W.~Chao, H.-k.~Guo and Y.~Zhang, \emph{{Majorana Dark matter with B+L gauge
  symmetry}}, \href{https://doi.org/10.1007/JHEP04(2017)034}{\emph{JHEP}
  {\bfseries 04} (2017) 034}
  [\href{https://arxiv.org/abs/1604.01771}{{\ttfamily 1604.01771}}].

\bibitem{He:1991qd}
X.-G.~He, G.C.~Joshi, H.~Lew and R.R.~Volkas, \emph{{Simplest Z-prime model}},
  \href{https://doi.org/10.1103/PhysRevD.44.2118}{\emph{Phys. Rev.} {\bfseries
  D44} (1991) 2118}.

\bibitem{Edsjo:1997bg}
J.~Edsjo and P.~Gondolo, \emph{{Neutralino relic density including
  coannihilations}},
  \href{https://doi.org/10.1103/PhysRevD.56.1879}{\emph{Phys. Rev.} {\bfseries
  D56} (1997) 1879} [\href{https://arxiv.org/abs/hep-ph/9704361}{{\ttfamily
  hep-ph/9704361}}].

\bibitem{Aprile:2017iyp}
{\scshape XENON} collaboration, \emph{{First Dark Matter Search Results from
  the XENON1T Experiment}},
  \href{https://doi.org/10.1103/PhysRevLett.119.181301}{\emph{Phys. Rev. Lett.}
  {\bfseries 119} (2017) 181301}
  [\href{https://arxiv.org/abs/1705.06655}{{\ttfamily 1705.06655}}].

\bibitem{Tan:2016zwf}
{\scshape PandaX-II} collaboration, \emph{{Dark Matter Results from First 98.7
  Days of Data from the PandaX-II Experiment}},
  \href{https://doi.org/10.1103/PhysRevLett.117.121303}{\emph{Phys. Rev. Lett.}
  {\bfseries 117} (2016) 121303}
  [\href{https://arxiv.org/abs/1607.07400}{{\ttfamily 1607.07400}}].

\bibitem{Patel:2015tea}
H.H.~Patel, \emph{{Package-X: A Mathematica package for the analytic
  calculation of one-loop integrals}},
  \href{https://doi.org/10.1016/j.cpc.2015.08.017}{\emph{Comput. Phys. Commun.}
  {\bfseries 197} (2015) 276}
  [\href{https://arxiv.org/abs/1503.01469}{{\ttfamily 1503.01469}}].

\bibitem{Patel:2016fam}
H.H.~Patel, \emph{{Package-X 2.0: A Mathematica package for the analytic
  calculation of one-loop integrals}},
  \href{https://doi.org/10.1016/j.cpc.2017.04.015}{\emph{Comput. Phys. Commun.}
  {\bfseries 218} (2017) 66}
  [\href{https://arxiv.org/abs/1612.00009}{{\ttfamily 1612.00009}}].

\bibitem{Belanger:2008sj}
G.~Belanger, F.~Boudjema, A.~Pukhov and A.~Semenov, \emph{{Dark matter direct
  detection rate in a generic model with micrOMEGAs 2.2}},
  \href{https://doi.org/10.1016/j.cpc.2008.11.019}{\emph{Comput. Phys. Commun.}
  {\bfseries 180} (2009) 747}
  [\href{https://arxiv.org/abs/0803.2360}{{\ttfamily 0803.2360}}].

\bibitem{Belanger:2018ccd}
G.~B\'elanger, F.~Boudjema, A.~Goudelis, A.~Pukhov and B.~Zaldivar,
  \emph{{micrOMEGAs5.0 : Freeze-in}},
  \href{https://doi.org/10.1016/j.cpc.2018.04.027}{\emph{Comput. Phys. Commun.}
  {\bfseries 231} (2018) 173}
  [\href{https://arxiv.org/abs/1801.03509}{{\ttfamily 1801.03509}}].

\bibitem{Pumplin:2002vw}
J.~Pumplin, D.~Stump, J.~Huston, H.~Lai, P.M.~Nadolsky and W.~Tung, \emph{{New
  generation of parton distributions with uncertainties from global QCD
  analysis}}, \href{https://doi.org/10.1088/1126-6708/2002/07/012}{\emph{JHEP}
  {\bfseries 07} (2002) 012}
  [\href{https://arxiv.org/abs/hep-ph/0201195}{{\ttfamily hep-ph/0201195}}].

\end{thebibliography}\endgroup
\end{document}